\documentclass[twocolumn,twocolappendix]{aastex631}

\usepackage{apjfonts,natbib,color,amsmath}
\usepackage{booktabs}
\usepackage{longtable}
\usepackage{xcolor}
\usepackage{xparse}

\usepackage{amsmath,tabularx} 

\usepackage{hyperref}

\NewDocumentCommand{\add}{+m}{%
 {\color{black}#1}%
}

\usepackage{tikz}
\usetikzlibrary{arrows.meta}
\usetikzlibrary{calc}
\usetikzlibrary{angles,quotes}
\usetikzlibrary{decorations.pathreplacing}

\usepackage{graphicx}
\newcommand{\beq}{\begin{equation}}
\newcommand{\beqa}{\begin{eqnarray}}
\newcommand{\eeq}{\end{equation}}
\newcommand{\eeqa}{\end{eqnarray}}
\newcommand{\ba}{\[\begin{aligned}}
\newcommand{\ea}{\end{aligned}\]}

\shorttitle{}
\shortauthors{}

\begin{document}

\title{
A TESS Test of the Hybrid Ring Strategy \add{for Technosignature Searches} Using GRB 221009A
\\
}


\author{Naoki Seto}
\affil{Department of Physics, Kyoto University, Kyoto 606-8502, Japan}


\begin{abstract}

We present the first observational test of the hybrid ring strategy, a general coordinated signaling scheme proposed by Seto (2025), which provides a practical  Schelling-point realization  for interstellar signaling. We use the exceptionally bright GRB 221009A as the anchoring flash for the scheme, together with the accurately measured distance to the Galactic center. This combination provides a high-precision relation linking sky position to a tightly constrained arrival-time window.
TESS observed the region around the GRB nearly continuously for $\sim$50 days in 2024, providing survey light curves that enable a direct test of this scheme with sharply predicted arrival-time windows of $\sim$3.4 days. Among 58 carefully selected stars, we identify two that show noticeable single-time-bin brightenings
 inside their predicted windows (where each time bin corresponds to a 200 s integrated TESS exposure). In both cases the brightenings coincide with excursions in at least one nearby star and are therefore most consistent with  instrumental origins. This test demonstrates that the hybrid ring strategy is practical with existing survey data and could serve as a promising basis for future technosignature searches. 

\end{abstract}






\section{Introduction}
\label{sec:intro}

Searches for technosignatures have so far yielded no confirmed detections, but this outcome does not by itself exclude the existence of extraterrestrial civilizations  \citep[e.g.][]{Drake1961,Tarter2001}.  
Such nondetections are consistent with the difficulty of exploring a search space spanning many independent dimensions, such as frequency, sky direction, distance, and signal morphology, in  particular for an intentional signal  \citep[e.g.][]{Wright2018b}.  
The resulting parameter space is enormous, and only a small fraction of it has been examined in practice.  
Because exhaustive coverage is infeasible, one of the major challenges in \add{Search for Extraterrestrial Intelligence} (SETI) is to identify principled ways to compress this space and make targeted searches more tractable. 
{A distant sender could reasonably anticipate this difficulty for any intentional signal and might therefore favor strategically designed transmission schemes that help a receiver limit the relevant search parameters.}

In game theory, a Schelling point is an implicit coordination outcome reached without prior communication  \citep{Schelling1960}.  
For interstellar signaling, conspicuous astrophysical transients can serve as coordination cues for independent civilizations.
These events could offer a geometrical  way to partially factor the vast search space, linking the sender’s position to a predicted arrival time without relying on any assumed signal morphology.
Early applications include the SETI ellipsoid, which uses light-travel geometry to relate transmission and reception epochs \citep[e.g.][]{1977Icar...32..464M,1980Icar...41..178M,1994Ap&SS.214..209L}  but still requires broad sky coverage and accurate distances to individual targets \citep[for recent studies, see e.g.][]{Davenport2022,Nilipour2023,Cabrales2024a}.  
The concurrent signaling scheme aims to remove the distance dependence and restrict both transmission and search directions to   rings on the sky \citep{Seto2019,Seto2025}.  
However, purely Galactic choices for the reference event face fundamental limitations, including insufficient distance accuracy and difficulty at specifying a base epoch \citep{Seto2019,Seto2020,Seto2024}.

The hybrid ring geometry provides a more robust realization of the signaling scheme \citep{Seto2025}.  
It  pairs a prominent extragalactic burst with the accurately known distance to the Galactic center \citep[e.g.][]{Gravity2021}.  
Because cosmological bursts lie at such large distances, their incoming wavefront is essentially planar across Galactic scales, so the burst distance does not enter into the geometric relation.  
In contrast, the Galactic center distance supplies the single well-measured length scale that determines the ring thickness,
\add{namely the finite angular width corresponding to the allowed arrival-time window at a given epoch.}
This combination yields exceptionally thin search rings on the sky for any fixed arrival time.

GRB\,221009A, dubbed the Brightest-Of-All-Time (BOAT; \citealt{Burns2023}), is the most luminous gamma-ray burst ever observed, exceeding all previous events in both peak flux and fluence.  
Such an exceptionally bright and rare event is expected to be  observed only once every ten thousand years and would likely be noticed independently by any technologically capable civilization. {Its sky location near the Galactic plane also increases the likelihood that the corresponding hybrid ring intersects many stars, offering a dense set of potential targets.}  
These properties make GRB\,221009A an outstanding reference event for applying the hybrid ring geometry.

The TESS mission \citep{Ricker2015} provides wide-field, continuous photometric
monitoring with near-uniform 200\,s exposures during its
$\sim27$-day observing sectors.
This coverage allows many stars on the predicted BOAT search ring to be examined near their predicted arrival times.  
Although TESS was not specifically designed for SETI, its long baselines and homogeneous cadence allow the geometric compression of the hybrid ring strategy to be exploited effectively.

Here we present the first observational implementation of the hybrid ring geometry.  
We analyze available TESS light curves of stars located  on the BOAT ring during the predicted arrival-time interval and search for short positive excursions detectable at the 200 s {exposure}.  
While our implementation uses spike-like brightenings as a practical observable, the hybrid framework itself is independent of any assumed signal morphology.  
{The methods developed here demonstrate that survey photometry can support geometry-based SETI strategies with sharply constrained spatial and temporal domains, and they provide an illustrative example for future observational programs.}

The remainder of this paper is organized as follows.
Section~2 summarizes the geometry of the hybrid ring produced by GRB~221009A,
including the associated $\pm3.4$\,day arrival window.
Section~3 describes the TESS observations of the BOAT region and the construction
of uniform, quality-controlled light curves.
Section~4 describes the definition and identification of
single-bin (200\,s)  flux excursions in the TESS light curves.
\add{Section~5 discusses the motivation for adopting these excursions as our primary
observational target and characterizes their statistical distribution across our light curve sample.}
Section~6 presents a detailed examination of the five highest-priority events,
including {bin}-level coincidence tests.
We summarize this study and discuss the implications for hybrid-ring SETI searches
in Section~7. \add{Appendix~A summarizes the properties of the final target sample.
Appendix~B presents an analytic estimate of the upper limits on the
transmitter power implied by the null result.
}

\section{Hybrid Geometry and Predictive Timing}
\label{sec:geometry}

This section summarizes the elements of the hybrid geometry that are required
for the observational analysis in this paper.  
The goal is to evaluate the arrival time mapping  associated with an extragalactic burst 
and to describe the resulting sky region to be analyzed.  
A schematic view of the geometry is shown in Fig.~\ref{figure:fig1}.

\subsection{Conceptual overview}

In the concurrent signaling scheme, multiple Galactic civilizations coordinate their intentional transmissions along a given light path so that the signals pass a common reference point at a common reference epoch.
The hybrid scheme generalizes this idea by incorporating information from a distant burst \citep[see][for a more detailed explanation including, for example,  the aberration effect]{Seto2025}.  
Specifically, it uses the wavefront of the burst at the moment it crosses the
Galactic center as a shared reference surface.  
When the burst is sufficiently distant, this wavefront can be treated as planar over Galactic scales.  
For a given intentional signal, the intersection of its light path with this
shared  surface defines both the reference point and reference   epoch.

We then apply this geometry to the case in which the recipient is the Sun.
Taking the Sun’s burst arrival time as the time origin, 
the time delay $\tau$ of any intentional signal is in 
one-to-one correspondence with the offset angle $\beta$ from the burst direction.  
This follows directly from the fact that differences in light-travel distance 
relative to the reference surface manifest as arrival time differences at the Sun.  
Consequently, all signals that arrive with the same delay $\tau$ must lie on
a narrow ring on the sky centered on the burst direction.

The relation between $\tau$ and $\beta$ takes a particularly simple form under
the plane-wave approximation.  
Its derivation is given in \citet{Seto2025}, and {here we summarize only the expressions 
relevant for the observational application below}.

\subsection{Time–angle relation}

\begin{figure}[t]
 \begin{center}
  \includegraphics[width=7.5cm,clip]{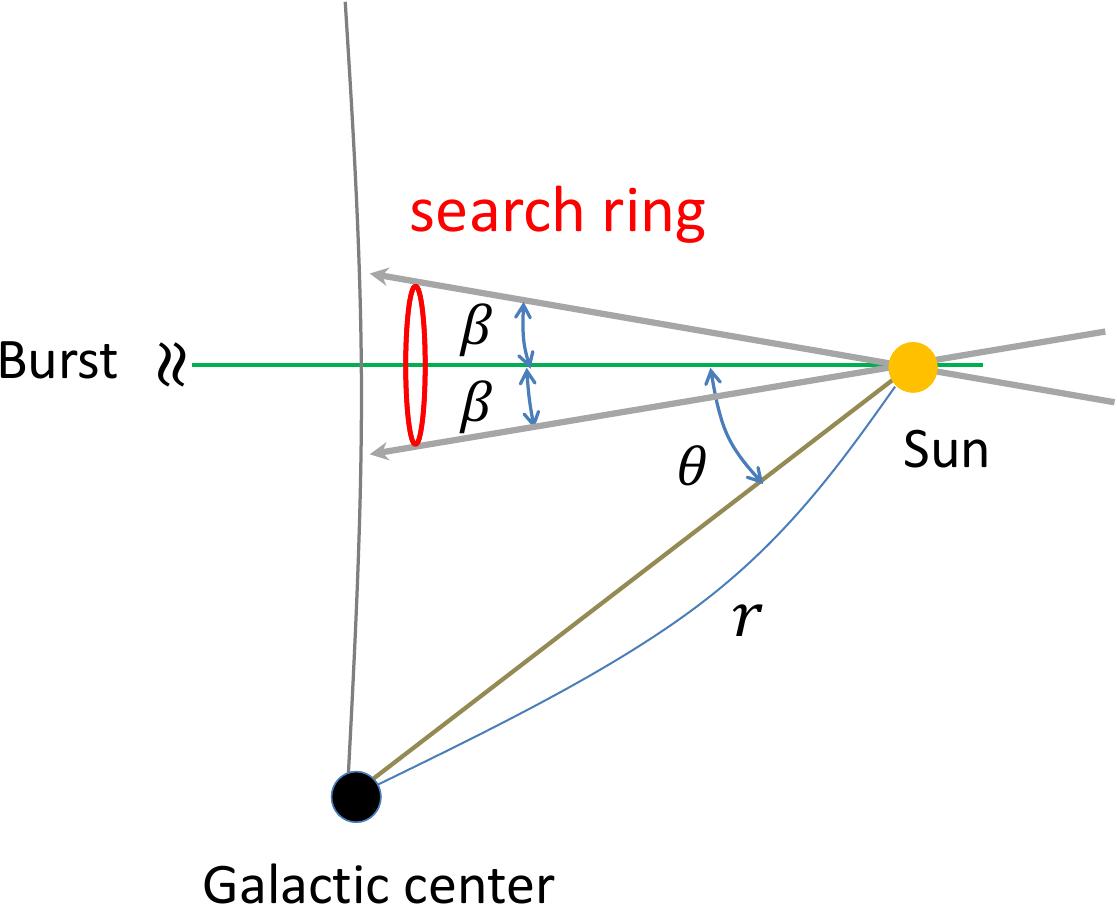}
\caption{
Geometry of the arrival-time ring used in this work.  
The angle between the Galactic Center and the reference burst (the BOAT in this paper), as viewed from the Sun, is denoted by $\theta$.  
In the hybrid scheme, the signal arrival-time delay $\tau$ after the burst is uniquely related to the ring opening angle $\beta$ around the burst direction.
}
  \label{figure:fig1}
 \end{center}
\end{figure}

Let $r$ denote the Sun–Galactic-center distance and let $\theta$ be the angular
separation between the burst and the Galactic Center, assuming $\theta<\pi/2$
(see Fig.~\ref{figure:fig1}).  
In the plane-wave limit appropriate for cosmological bursts, the delay $\tau$
and the ring angle $\beta$ satisfy  
\begin{equation}
    \tau = \frac{r}{c} \cos\theta\,(\sec\beta - 1)\,.
    \label{eq:tau_basic}
\end{equation}
This expression provides a one-to-one mapping between the delay $\tau$ and the
opening angle $\beta$ of the arrival-time ring.  
Solving for $\beta$ gives  
\begin{equation}
    \beta(\tau) =
    \arccos\!\left[
        \frac{r\cos\theta}{c\tau + r\cos\theta}
    \right]\,.
    \label{eq:beta_of_tau}
\end{equation}

\add{For completeness, we note that the hybrid scheme implies a causal depth
$r\cos\theta$, defined as the maximum line-of-sight distance over which a
responding signal can remain causally connected to the reference burst,
given the finite speed of light \citep{Seto2025}.}

\subsection{Dependence on the Galactic-center distance}

The Sun–Galactic-center distance provides the only physical scale in
Eq.~(\ref{eq:tau_basic}).  
We adopt  
\[
    r = 8275~\mathrm{pc}
\]
from recent infrared astrometry \citep{Gravity2021}.  
Since the mapping from $\tau$ to $\beta$ scales linearly with $r$, 
the fractional uncertainty $\Delta r / r \simeq 0.005$ directly determines 
the uncertainty in the predicted arrival time delay.  
Differentiating Eq.~(\ref{eq:tau_basic}) gives  
\begin{equation}
    \Delta\tau \simeq \tau\left(\frac{\Delta r}{r}\right)\,.
\end{equation} 
At  $c\tau\ll r\cos\theta$, the corresponding angular width of the arrival-time ring is  
\begin{equation}
    \Delta\beta \simeq \frac{\beta}{2}\left(\frac{\Delta r}{r}\right)\,.
\end{equation}

\subsection{Application to GRB 221009A}

The BOAT GRB has direction  
\[
    \mathrm{RA} = 288.2646^{\circ}, \qquad
    \mathrm{Dec} = 19.7734^{\circ},
\]
which gives an angular separation from the Galactic Center of  
\[
    \theta \simeq 53.1426^{\circ}\,
\]
with the associated causal depth $r\cos\theta=4.964$\,kpc.

Throughout this work we use the barycentric TESS Julian Date system,
\[
    \mathrm{BTJD} \equiv \mathrm{BJD} - 2457000\,,
\]
which is the native time coordinate of TESS data products.  
The burst arrival time is  
\[
    t_{\mathrm{burst, BTJD}} = 2862.0548\, .
\]
Thus, for a given opening angle $\beta$ around the BOAT, the predicted arrival time $t_{\rm obs}$ is given as
\beq
t_{\rm obs}=t_{\mathrm{burst, BTJD}}+\tau. \label{tobs}
\eeq

TESS observed the BOAT region in Sectors 80 and 81 (explained later in more detail).  
The final observation time in Sector 81 was  
\[
    t_{\mathrm{S81, end}} \simeq 3533.18\,,
\]
so the elapsed time between the burst and the end of our coverage was  
\[
    \Delta T \equiv
    t_{\mathrm{S81, end}} - t_{\mathrm{burst, BTJD}}
    \simeq 6.7\times10^{2}\ \mathrm{days}\,.
\]

Using the fractional distance uncertainty $\Delta r / r \simeq 0.005$ 
reported in \cite{Gravity2021}, we obtain  
\[
    \Delta\tau \simeq \Delta T\,(\Delta r / r) \simeq 3.35\ \mathrm{days}\,.
\]

{Motivated by this estimate, we adopt a symmetric observational window of}
\[
    \Delta\tau_{\mathrm{win}} = \pm 3.4\ \mathrm{days}\,.
\]

This completes the construction of the arrival-time window and the associated
arrival-time ring.  
The application to TESS observations is described in Section~3.

\section{TESS Data and Sample Selection}
\label{sec:data}

In this section we describe how the post-BOAT TESS observations were used to
assemble the stellar sample for our light curve analysis.
Using the hybrid-geometry prediction of Section~2, we first select cataloged
stars that fall on the arrival-time ring during 
Sectors~80 and~81.

This geometric pre-selection yields 1035 candidates.
We then apply a uniform sequence of filters: availability of public TESS light
curves, conservative Gaia~DR3 quality cuts, a neighbor screen to remove
blended or contaminated apertures, and a brief light-curve check.
After these steps, the working sample consists of 58 stars, which form the
basis for the spike statistics and case studies in the later sections.

\subsection{Overview of Post-BOAT TESS Coverage}

Any coordinated signal associated with GRB\,221009A must arrive after the
burst, and our search is therefore restricted to post-burst TESS data.
The BOAT direction lies at a low ecliptic latitude, and TESS revisited this
region only twice after the 2022 event, in Sectors~80 and~81 in 2024.

Sector~80 spanned 2024 June 18 to July 14
($\mathrm{BTJD}$: $3479.9$--$3506.3$), and Sector~81 provided a consecutive
visit from 2024 July 14 to August 9 ($\mathrm{BTJD}$: $3506.5$--$3533.2$),
yielding a continuous post-BOAT baseline of $53.4$\,days.

As discussed in Section~2, the hybrid geometry predicts a unique
arrival-time center $t_{\rm obs}$ for each star, with an uncertainty of
approximately $\pm 3.4$\,days arising from the fractional error in the
Sun–Galactic-center distance. To ensure that the entire $\pm 3.4$\,day
window is covered by TESS, we require
\begin{equation}
    3487 \le t_{\rm obs} \le 3526,
    \label{eq:tobs_window}
\end{equation}
which lies safely inside the combined temporal span of Sectors~80 and~81.
This timing requirement therefore defines the subset of ring-selected stars
for which the TESS data permit a complete test of the hybrid prediction.

\subsection{From arrival-time center to ring geometry and initial catalog}

The hybrid scheme establishes a one-to-one mapping between the
arrival-time delay $\tau$ and the angular offset $\beta$ from the burst
direction (Section~2).  Using Eqs.~ (\ref{eq:beta_of_tau}) and (\ref{tobs}), the allowed range of
$t_{\rm obs}$ in Eq.~(\ref{eq:tobs_window}) maps to a narrow interval of
opening angles,
\begin{equation}
    0.8330^\circ \le \beta \le 0.8585^\circ ,
    \label{hanni}
\end{equation}
corresponding to a ring width of roughly $0.03^\circ$ around the burst
direction.  In what follows, we refer to this BOAT-centered annulus as
the arrival-time ring.

We then selected all TESS Input Catalog sources located on  this annulus
during Sectors~80 and~81.  A simple magnitude cut,
\begin{equation}
   8.0 \le T_{\rm mag} \le 15.5 ,
   \label{lum}
\end{equation}
was applied to avoid saturation and to exclude stars too faint for
reliable spike detection.  These geometric and photometric criteria yield
1035 sources on the arrival-time ring, with the brightest one having $T_{\rm mag}=9.96$.

The annulus subtends only $\sim0.1~\mathrm{deg}^2$, compared with the
$24^\circ\times96^\circ$ field of view of a single TESS pointing
\citep{Ricker2015}.  The hybrid construction therefore compresses the
directional search space by about four orders of magnitude, a key
advantage for survey-based technosignature searches.

\subsection{Photometric coverage and QLP requirement}
To carry out a  uniform spike search along the BOAT ring, we use
publicly available light curves for the ring-selected stars.  
In practice,
this corresponds to the products of the TESS Quick Look Pipeline (QLP;
\citealt{Huang2020}), which provide aperture photometry for bright stars in
the full-frame images.  
Within the narrow annulus defined in Section~3.2, the QLP products are the
only consistently processed light curves available for stars in the
arrival-time region.

From the 1035 catalog entries on the BOAT ring, we next identify those with
QLP light curves in both Sector~80 and Sector~81.  
This condition is not
required by the hybrid geometry, but it ensures that each target is
observed across the full post-BOAT temporal baseline available in the
archive and allows direct checks for sector-dependent artifacts.  
Applying this criterion leaves 164 stars with QLP coverage in both sectors.

Higher-cadence light curves produced by the Science Processing Operations
Center (SPOC; \citealt{Jenkins2016}) exist for a small number of preselected
targets in the surrounding field.  Unfortunately, 
none of these targets lie inside the
arrival-time annulus, and therefore the SPOC products do not contribute to
our post-BOAT sample.

\subsection{Gaia quality filters}

For the 164 stars with QLP light curves in both sectors, we applied a small set
of conservative Gaia DR3 quality requirements to exclude objects affected by
blending, inconsistent photometry, or known strong variability \citep{Gaia2022}.
The following four Gaia DR3 criteria were adopted:
\begin{enumerate}
    \item \texttt{ruwe < 1.35}, removing objects with unreliable or
          non-single-star astrometric solutions;
    \item \texttt{phot\_bp\_rp\_excess\_factor < 1.5}, rejecting sources with
          inconsistent BP/RP fluxes;
    \item \texttt{duplicated\_source = False}, ensuring a unique Gaia
          cross-identification;
    \item \texttt{phot\_variable\_flag $\neq$ `VARIABLE'}, excluding
          large-amplitude variables that could mimic spike-like behavior.
\end{enumerate}

These requirements reduce the sample to 108 stars.

Proper-motion corrections were applied after the Gaia filtering when
computing the predicted arrival-time centers.
Although the corrections are small \add{(with median shifts of $\sim 0.008$ arcsec in angular position and $\sim 0.004$ days in arrival time, and maxima well below 0.1 arcsec and 0.05 days across the sample)}, propagating each source to the
mid-epoch of the TESS observations (2024 July 14) provides a consistent
temporal reference for the window selection.

\subsection{Neighbor screening}

TESS full-frame images have a large pixel scale (21\arcsec\ per pixel), and
the point-response function (PRF) has extended wings that distribute flux over
multiple pixels.  
Bright neighbors can therefore contaminate the QLP apertures
and produce spurious  features \citep[e.g.][]{Huang2020}.  
To remove
such cases we applied a two-stage neighbor screen using Gaia DR3 positions and
magnitudes.

Starting from the 108 stars that passed the Gaia filters, we applied the following two-stage procedure:

\begin{enumerate}
    \item \textbf{Primary exclusion (within 21\arcsec):}
    Any target with a Gaia neighbor closer than 21\arcsec\ and brighter by
    $\Delta G < 2.5$ mag was removed, since the PRF wings can contribute a
    non-negligible fraction of the neighbor's flux to the QLP aperture.

    \item \textbf{Secondary classification (21--42\arcsec):}
    For neighbors in the 21--42\arcsec\ annulus, again requiring
    $\Delta G < 2.5$ mag, the expected contamination is modest.  
    Targets with
    such neighbors were retained but flagged as \add{``R-type'' (\textit{restricted}),
while those without were labeled ``P-type'' (\textit{pristine}).}
\end{enumerate}

After applying the primary 21\arcsec\ exclusion, the Gaia-filtered set of
108 stars was reduced to 60 stars, including 39 R-type objects.

\subsection{Final data-quality checks}

The 60 stars that passed the neighbor screening were examined for
data-quality issues in their QLP light curves from Sectors~80 and~81.
{Time bins} with \texttt{QUALITY = 0} define the reliable baseline for spike
searches, yet two objects (TIC~353166070 and TIC~353166074) show substantial
intervals of missing data even within these {time bins} and were therefore
removed.

No similar issues were found in the other targets.
The final working sample contains 58 stars with complete post-BOAT QLP
coverage in both sectors and serves as the input to the spike analysis in
Section~4.

\add{For clarity, we briefly summarize the terminology related to the 200 s temporal
sampling used throughout this work.
An \emph{exposure} refers to a single 200 s TESS full-frame measurement.
We use the term \emph{time bin} when referring to its role as a discrete
statistical unit in the spike analysis, and \emph{integration} when emphasizing
the flux accumulated over this interval.
}

\subsection{Photometric and astrometric properties of the final sample}

\add{Table~\ref{tab:targets_major}, provided in Appendix~A, summarizes the basic properties of the
58 stars that passed all selection steps.}

After applying the QLP availability, Gaia DR3 quality filters, and the
neighbor screen, the resulting final sample occupies a narrower  magnitude range of $T_{\rm mag} = 10.0$ to 13.4, in contrast to
the range in Eq.~(\ref{lum}) for the initial 1035 stars.

Gaia DR3 parallaxes for the 58 stars typically lie between 0.1 and 2~mas,
corresponding to approximate distances of 0.5 to 10~kpc.  
Formal uncertainties are generally $\sim 0.01$ to $0.02$~mas.  
A few objects have very small parallaxes ($<0.1$~mas), implying large
fractional errors and poorly constrained distances.  
Such stars are likely to lie beyond the causal depth of
$\sim 5~\mathrm{kpc}$ associated with the BOAT (see Section~2).

\subsection{On-sky distribution and CCD-footprint constraints}

\begin{figure}[t]
 \begin{center}
  \includegraphics[width=11.5cm,clip]{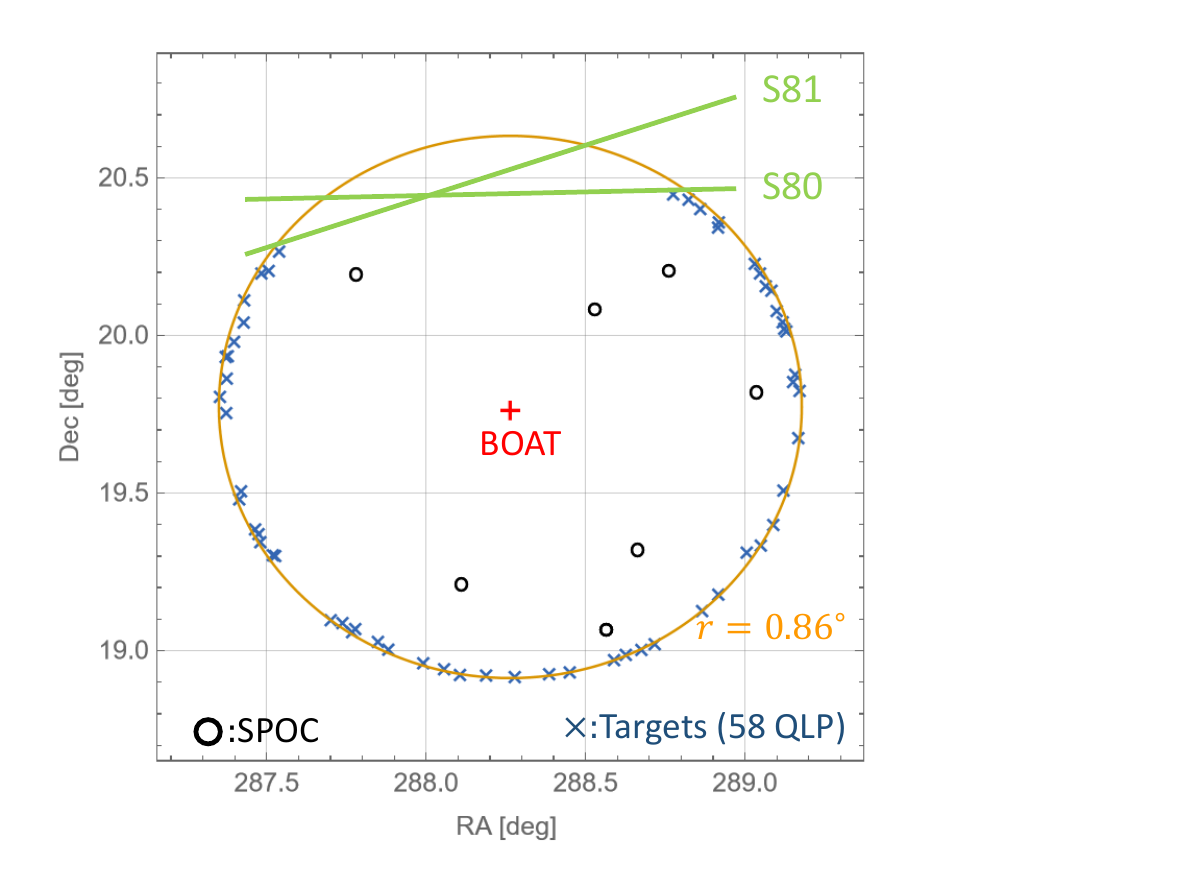}
 \caption{
Sky distribution of the final 58 QLP targets (blue crosses) around the
BOAT direction.  
The stars trace the thin annulus defined by 
$0.833^\circ \le \beta \le 0.858^\circ$ (Section~2).  
Green lines mark the boundaries of the TESS footprints for
Sector~80 (Camera~2, CCD\,4) and Sector~81 (Camera~2, CCD\,3). 
Only the region where these two footprints overlap provides QLP coverage in
both sectors, producing the truncated morphology of the sample.  
Seven SPOC targets within $0.86^\circ$ of the BOAT direction are shown as
black circles; none lie inside the arrival-time annulus.
}
  \label{figure:fig2}
 \end{center}
\end{figure}

Figure~\ref{figure:fig2} shows the sky distribution of the final 58
QLP targets.  
A characteristic feature of this distribution is the imprint of the
Sector~80 and Sector~81 CCD footprints.  
During these visits the BOAT field
fell on Camera~2, CCD\,4 (Sector~80) and Camera~2, CCD\,3 (Sector~81).  
Only the region where these two footprints overlap provides QLP coverage in
both sectors, so stars outside this overlap were excluded as described in Section~3.3.  
This geometric constraint naturally produces the truncated morphology seen
in Figure~\ref{figure:fig2}.

A small subset of stars are located very close to the CCD boundaries.  
For Sector~80 the closest objects are TIC~353516003 (the nearest), followed
by TIC~353516889 and TIC~354059270.  
For Sector~81 the corresponding closest stars are TIC~9641063 (the nearest),
TIC~9640619, and TIC~9640566.

Note also that during Sectors~80 and~81 seven SPOC targets lie within
$0.86^\circ$ of the BOAT direction, but none fall inside the arrival-time
annulus.

\begin{table}[t]
\centering
\caption{
Common observing segments used in the analysis, defined from
\texttt{QUALITY = 0} {bins} in Sectors~80 and~81.
Times are expressed in BTJD $\equiv \mathrm{BJD}-2457000$.
Only intervals longer than 5~h are included.
The final column lists the corresponding TESS sector.
}
\label{tab:common_segments}
\begin{tabular}{rcccc}
\hline
Index & BTJD$_{\rm start}$ & BTJD$_{\rm end}$ & Duration [h] & Sector\\
\hline
1  & 3479.958 & 3481.324 & 32.78 & 80\\
2  & 3481.352 & 3481.755 & 9.66  & 80\\
3  & 3481.766 & 3484.581 & 67.56 & 80\\
4  & 3489.056 & 3491.403 & 56.33 & 80\\
5  & 3491.412 & 3491.831 & 10.05 & 80\\
6  & 3491.954 & 3492.371 & 10.00 & 80\\
7  & 3493.602 & 3499.998 & 153.50 & 80\\
8  & 3500.213 & 3504.114 & 93.61 & 80\\
9  & 3504.123 & 3505.878 & 42.11 & 80\\
10 & 3505.887 & 3506.292 & 9.72  & 80\\
11 & 3506.590 & 3509.446 & 68.55 & 81\\
12 & 3509.455 & 3513.383 & 94.27 & 81\\
13 & 3516.860 & 3518.237 & 33.05 & 81\\
14 & 3518.247 & 3518.524 & 6.66  & 81\\
15 & 3518.534 & 3518.779 & 5.89  & 81\\
16 & 3519.865 & 3526.105 & 149.77 & 81\\
17 & 3526.332 & 3530.765 & 106.38 & 81\\
18 & 3532.284 & 3533.091 & 19.39 & 81\\
\hline
\end{tabular}
\end{table}

\section{Preparation of the TESS Light Curves}
\label{sec:targets}
In this section we process the TESS light curves of the 58 targets for the
spike analysis.  We first define the usable \texttt{QUALITY = 0} {bins} and
then construct a standardized time-series representation of each target.
These steps provide a uniform basis for identifying short-duration
excursions in Section~5.

\subsection{Identification of Usable \texttt{QUALITY = 0} Intervals}
\label{sec:quality0_selection}

For each of the 58 targets we begin by selecting the \texttt{QUALITY = 0}
{bins} in the QLP full-frame products.
To ensure stable local statistics, three uniform time-domain filters are
applied to every light curve:  
(1) {time bins} within 15 minutes of each sector boundary are removed;  
(2) one 200\,s {bin} is trimmed from each end of the remaining intervals;  
(3) only intervals longer than 5\,h are retained.  
These conservative rules remove only a small fraction of otherwise usable data.

Applying the same filters to all 58 targets produces a small
set of clean intervals.  
We then take the intersection of these per-target
intervals to define the portions of the TESS time series that are usable in
common across all 58 stars.

This procedure yields 18 contiguous observing segments, each sampled at the
native 200\,s {bin}, within the Sectors~80 and~81 window
(BTJD~3480.0--3533.1), providing 40.4\,days of usable coverage
(a duty cycle of about 76 percent).

\subsection{Light Curve Preparation and Standardization}
\label{sec:detrend}

A slowly varying background on half-day or longer time scales is present
in many QLP light curves [denoted here as $f(t)$] and can obscure short-duration excursions when
the data are viewed over the full 50\,day baseline.  
To suppress this slow background while preserving the statistics of brief
spikes, we apply a mild detrending using a 12\,h rolling-median filter
(see Figures~\ref{fig:spike_lc_primary} and \ref{fig:spike_lc_summary}).  
{As shown in Section~6, this detrending has only a negligible effect on
the standardized flux values, ensuring that it does not bias the spike
statistics.}

This procedure defines the detrended flux
\[
    f_{\rm det}(t) = f(t) - \mathrm{median}_{12{\rm h}}[f(t)],
\]
which is used only to improve the visual clarity of the full multi-day
light-curve plots.

Within each of the 18 common observing segments, we then compute a rolling
median $\mu_{\rm loc}(t)$ and a rolling robust scatter
$\sigma_{\rm robust}(t)$ using a 4\,h window.
The scatter is defined from the median absolute deviation (MAD) as
\begin{equation} \label{sigm}
    \sigma_{\rm robust}(t)
    = 1.4826 \times \mathrm{MAD}\!\left(f_{\rm det}(t)\right),
\end{equation}
where the factor $1.4826$ converts MAD into an equivalent Gaussian
standard-deviation scale.

{This convention allows a common scale to be used across targets and segments 
without assuming Gaussian noise.}
A floor equal to the 20th percentile of $\sigma_{\rm robust}$ in each
segment is applied to stabilize the estimator.

The standardized flux used throughout the spike search is then
\begin{equation}
    z(t)
    = \frac{ f_{\rm det}(t) - \mu_{\rm loc}(t) }
           { \sigma_{\rm robust}(t) } ,
    \label{eq:zscore}
\end{equation}
which provides a uniform measure of short-duration excursions across all
targets.
This time series  $z(t)$ forms the statistical basis for the {time-bin}-level
analysis in Section~4.3 and the spike counts discussed in Section~5.

\subsection{Segment-Level Quality Assessment}
\label{sec:seg_quality}

Although all 18 consensus segments contain only {bins} marked
\texttt{QUALITY = 0}, the noise properties of the rolling robust scatter
$\sigma_{\rm robust}(t)$ vary across segments and targets.
We quantify these variations using two diagnostics that summarize the
segment-level behavior of the local noise.

The first diagnostic is a robust coefficient of variation,
\[
\mathrm{CV}_{\rm robust}
 = \frac{1.4826\,\mathrm{MAD}\!\left(\sigma_{\rm robust}(t)\right)}
        {\mathrm{median}\!\left(\sigma_{\rm robust}(t)\right)} .
\]

The second diagnostic is a percentile ratio of the scatter $\sigma_{\rm robust}(t)$ in each  segment,
\[
\mathcal{R}_{90/10}
 = \frac{\sigma_{\rm p90}}{\sigma_{\rm p10}} .
\]

Across the full set of $58\times18 = 1044$ target–segment combinations,
twenty satisfy either
$\mathrm{CV}_{\rm robust}\gtrsim 0.3$ or
$\mathcal{R}_{90/10}\gtrsim 2$.
These represent only about two percent of all combinations. {We refer to these as bad segments.}
Because the affected fraction is small and the diagnostics are heuristic,
we retain all segments for the subsequent analysis.
{Although these metrics are not used as vetoes, we note that the
segments flagged by these criteria do tend to show noisier behavior when
examined in detail.  The diagnostics therefore serve mainly to highlight
potentially unstable intervals rather than to exclude data.}

\subsection{Consolidated {time-bin}-level Data Set}
\label{sec:prep_spike}

For the statistical analysis we use the standardized flux $z(t)$
in the native 200\,s time bins of the QLP light curves,
restricting the analysis to the common observing segments listed in Table 1.

To relate each {time bin} to the expected signal timing, we assign an
in-window label based on the predicted arrival time $t_{\rm obs}$ from
Section~2:
\[
    |t - t_{\rm obs}| \le 3.4~\mathrm{day}.
\]
This timing flag is used later to separate spike counts inside and
outside the predicted time window.

\subsection{Summary}

Section~4 establishes a uniform time-series basis for all 58 targets.
The usable data are defined by the 18 consensus segments composed of
\texttt{QUALITY = 0} {bins} in Sectors~80 and~81.
Within these segments each light curve is characterized through the
standardized flux $z(t)$ derived from local robust statistics, and every
{bin} is assigned an in-window or out-of-window timing label.
This consolidated data set serves as the direct input for the spike
analysis in Section~5.

\section{Spike Statistics}

\add{In this section, we characterize spike events in the TESS light curves
across the BOAT-ring target sample.
}

\subsection{Spike-like Excursions as an Observational Representation}
\add{
Before presenting the spike statistics, we briefly explain why
single-exposure (200 s) brightenings
are adopted as the basic observational unit in this work.

In the TESS full-frame images, flux is integrated over 200\,s, which defines the
natural temporal unit of the QLP light curves used in this analysis. Any
sufficiently short artificial optical  pulse would therefore appear
as a positive single-bin excursion in the QLP light curves. The
spike-like morphology considered here thus reflects temporal integration by the
instrument, rather than an assumption about the intrinsic emission timescale or
signaling strategy of a putative transmitter.

Both astrophysical variability and instrumental systematics can, in
principle, generate isolated 200\,s excursions in survey photometry.
Discriminating between these possibilities is therefore essential.
In this work, we discuss this issue in Section~6.1 using dedicated
event-level diagnostics.
}

\subsection{Overview of the 58-Target Sample}
\label{sec:overview_spikes}

The combined light curves contain roughly $10^{6}$ independent  200 s time bins.  
We begin by characterizing the basic population of single-bin excursions, which sets the statistical background for the analyses that
follow.

We classify standardized flux excursions using two thresholds:
\emph{soft} spikes for $3.5 \le |z| < 5$ and \emph{hard} spikes for
$|z| \ge 5$.  
These thresholds serve only as empirical markers of the TESS noise
environment and do not reflect any assumed physical pulse amplitude.

\add{Across the sample, soft spikes are common: most targets show 10--25 events,
although the full range extends from one (TIC~354062527) to more than one hundred
(TIC~353165889). 
Hard positive spikes are rare: only ten targets show any event with
$z\ge 5$, and only four reach $z\ge 6$.}

To evaluate short-timescale coherence, we examined consecutive excursions.
Across all targets, only nine instances of adjacent {time bins} satisfy
$|z|\ge 3.5$, and just one sequence contains two consecutive events with
$|z|>5$ ($z=-6.46$ followed by $-5.42$  around BTJD 3494.4 in TIC~353165889).  
The scarcity of such sequences indicates that spike amplitudes are nearly
uncorrelated on 200\,s timescales, supporting the use of single-{bin}
excursions as the fundamental units in the following subsections.

Finally, the statistical distribution identifies the specific events
selected for detailed inspection in Section~6: two targets
(TIC~354057959 and TIC~353165889 presented in Fig. \ref{fig:spike_lc_primary}) host hard positive spikes within the
$\pm3.4$\,day BOAT arrival-time window, and three others
(TIC~10121249, TIC~9640566, and TIC~10121399 in  Fig. \ref{fig:spike_lc_summary}) exhibit the most extreme
positive excursions in the sample with $z\ge 10$, outside the arrival-time window.  In the next section we present a detailed examination of these five light curves. \add{In Fig.~\ref{fig:spike_lc_primary}, the bottom panel (TIC~353785997) is included for comparison with the middle panel and is not part of the 58 selected targets.}

\subsection{Distribution of Spike Amplitudes}
\label{sec:global_dist}

With approximately $10^{6}$ {time bins} across the sample, the Gaussian
expectation for large excursions is extremely small:
\begin{align}
N_{\rm exp}(|z|\ge3.5) &\approx 470 , \\
N_{\rm exp}(z\ge5)     &\approx 0.3 , \\
N_{\rm exp}(z\ge6)     &\approx 0.001 .
\end{align}
The observed population is far larger: more than ten targets show at least
one $z\ge5$ event, and several reach $z\ge6$. 
This demonstrates a strong heavy-tailed component in the TESS full-frame
light curves, consistent with earlier reports of non-Gaussian statistics in
space-based photometry \citep[e.g.,][]{Christiansen2012}.

\subsection{Segment-level Context}

To gauge how noise variations influence the spike statistics, we examined
the 18 shared observing segments using the robust scatter diagnostics
introduced in Section~4.3.  These diagnostics are descriptive only and are
not used to exclude data.

\add{Bad segments constitute roughly two percent of the usable timeline (see
Section~4.3), yet contribute 10--50 percent of all spikes, with a particularly
strong excess of negative excursions.
  This indicates that part of the negative tail is
sensitive to episodic detector noise or background fluctuations.}

In contrast, neither of the two in-window positive hard spikes occurs in a
bad segment (see Fig. 3).  Their timing is therefore not trivially linked to intervals
of degraded noise conditions.

\begin{figure*}[t]
  \centering
   \includegraphics[width=1\textwidth]{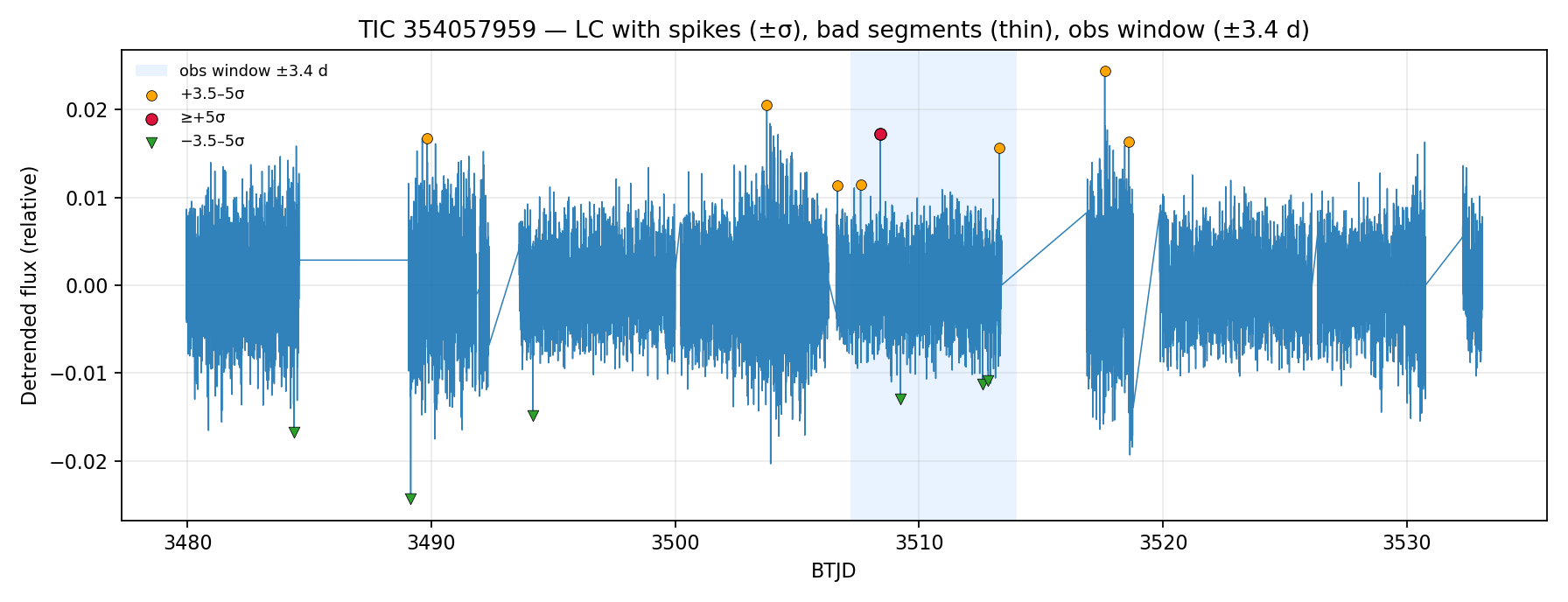}
   \includegraphics[width=1\textwidth]{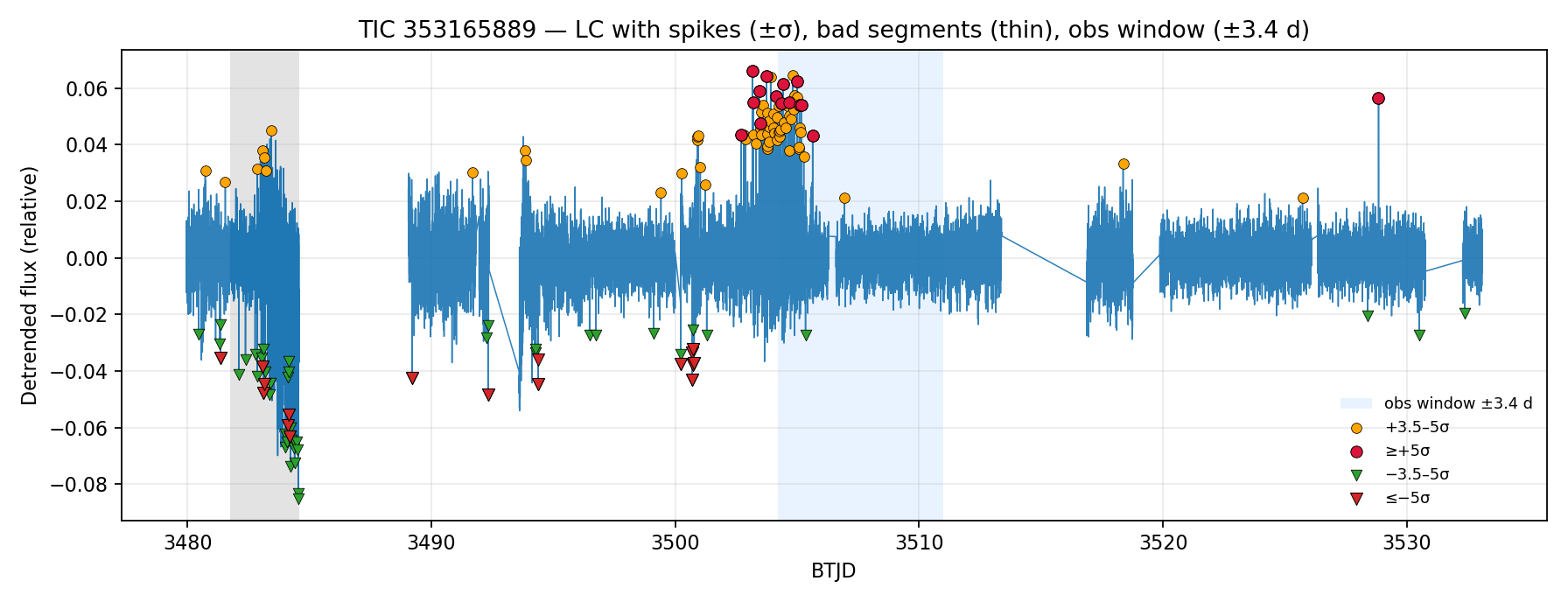}
   \includegraphics[width=1\textwidth]{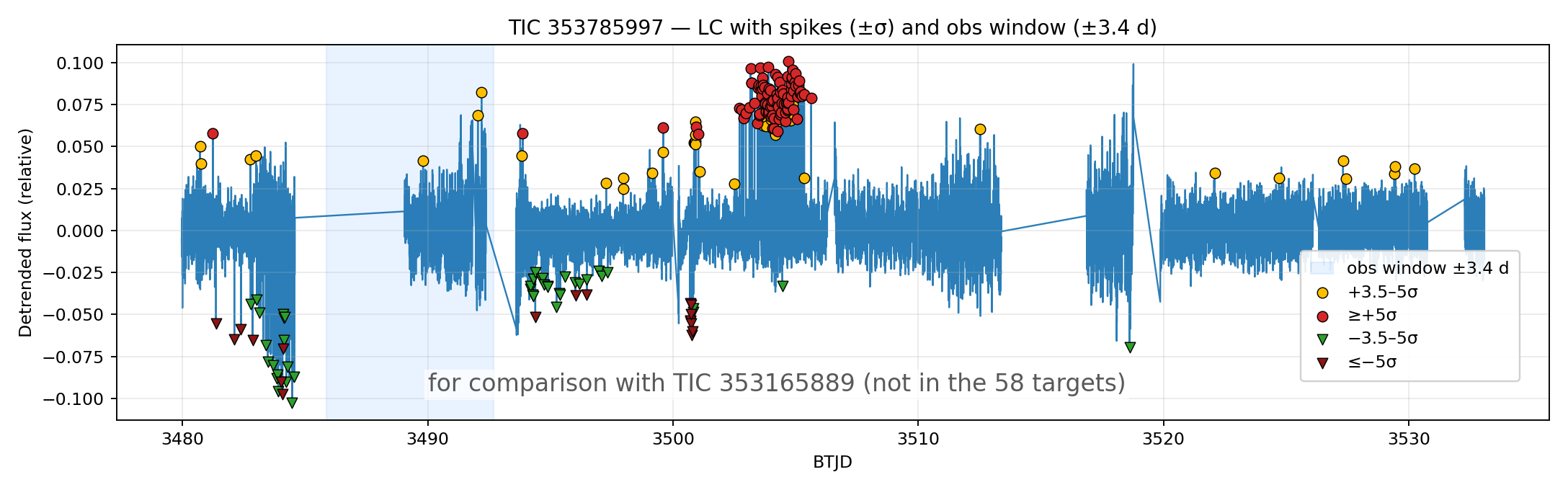}
\caption{
Detrended QLP light curves for (top) TIC~354057959, (middle) TIC~353165889,
and (bottom) TIC~353785997.
Blue shading indicates the $\pm 3.4$\,day BOAT search windows, while light gray
shading marks bad segments {(not excluded from the spike statistics, see Sec. 4.3)}.
Hard spikes ($z\ge5$) are shown in red, soft spikes ($3.5\le z<5$) in orange,
and negative excursions in green ($-5<z\le-3.5$) and dark red ($z\le-5$).
\add{TIC~353785997 lies about $771\arcsec$ from TIC~353165889 and is not among the
58 ring-selected targets; it is shown here solely for comparison.
The two stars exhibit nearly simultaneous positive excursions near BTJD~3503.}
}
  \label{fig:spike_lc_primary}
\end{figure*}

\begin{figure*}[t]
  \centering
  \includegraphics[width=1\textwidth]{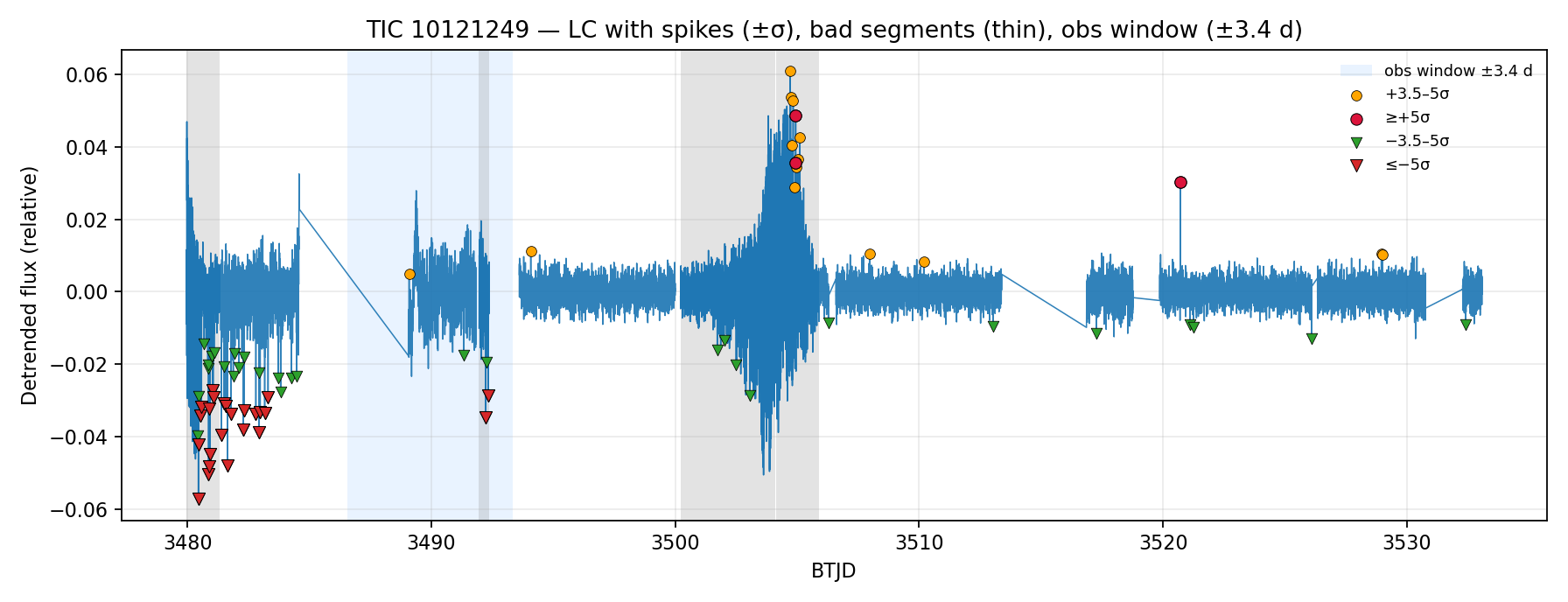}
  \includegraphics[width=1\textwidth]{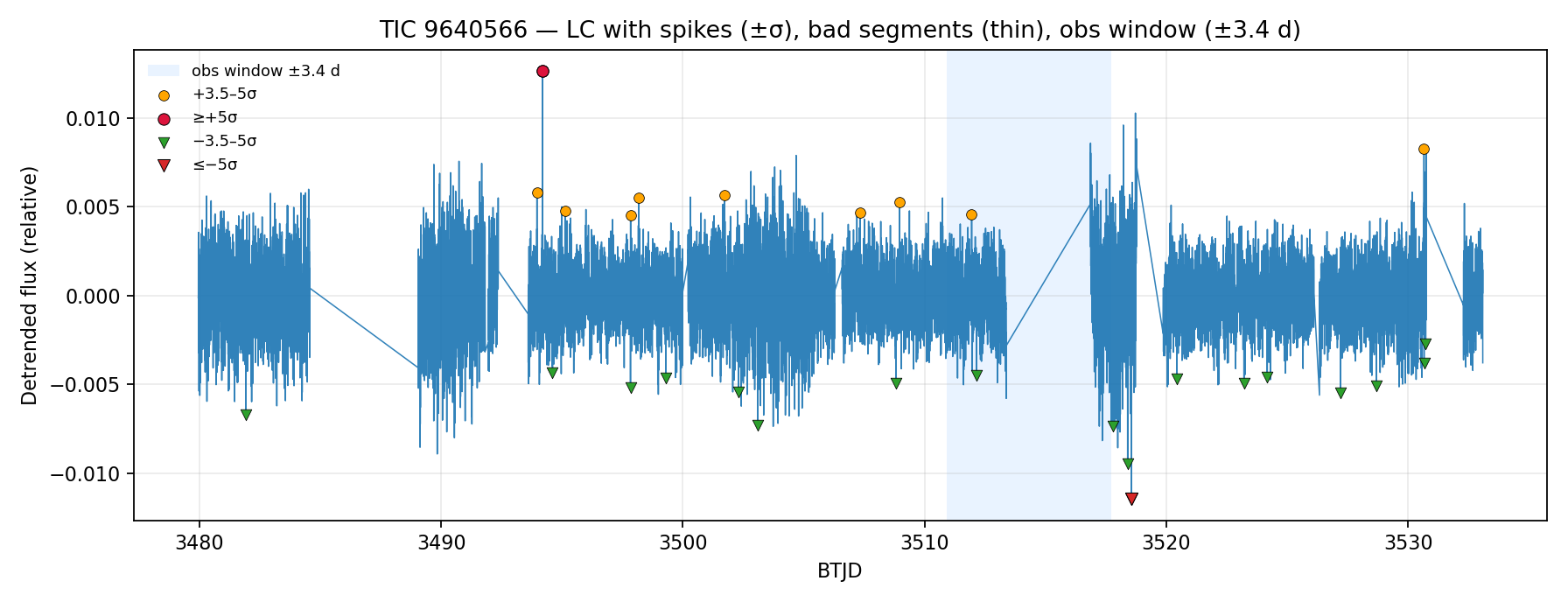}
  \includegraphics[width=1\textwidth]{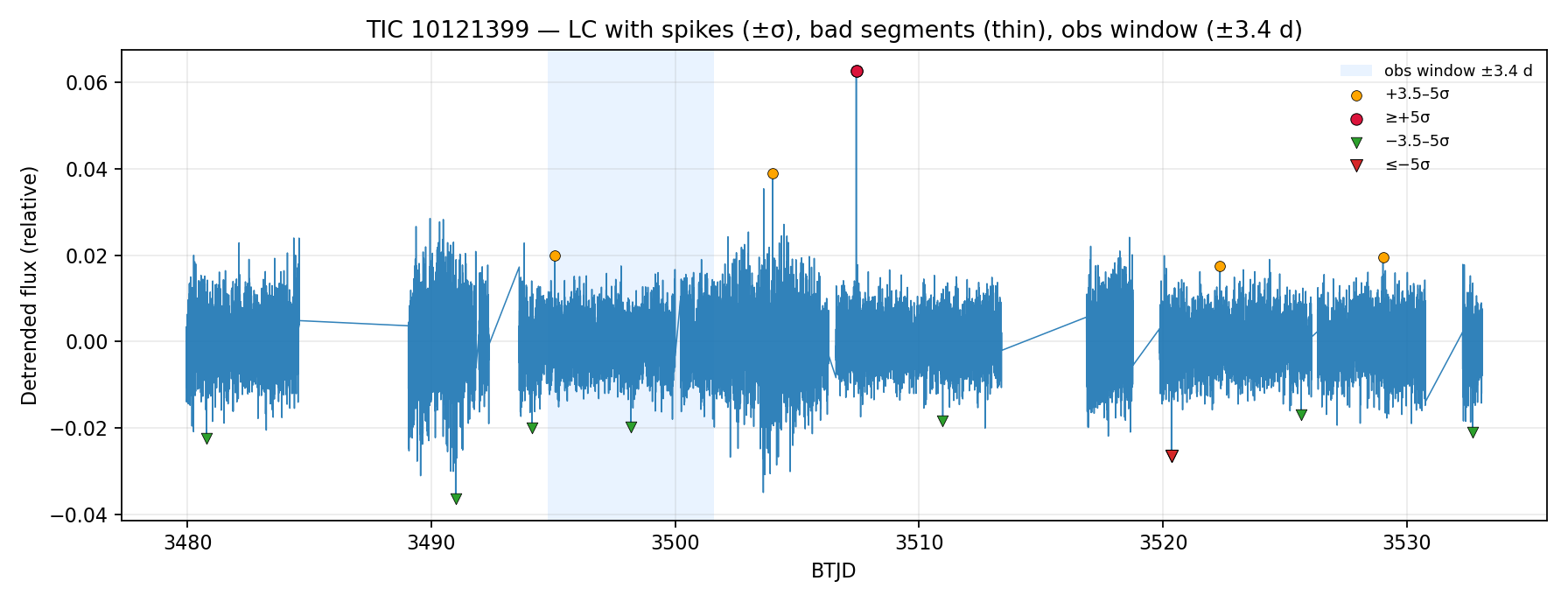}
  \caption{
Detrended QLP light curves for the three targets exhibiting the largest
positive excursions in our sample ($z\ge 10$):
(top) TIC~10121249, (middle) TIC~9640566, and (bottom) TIC~10121399. The three spikes are outside the arrival time windows. 
  }
  \label{fig:spike_lc_summary}
\end{figure*}

\begin{figure}[t]
 \begin{center}
  \includegraphics[width=8.5cm,clip]{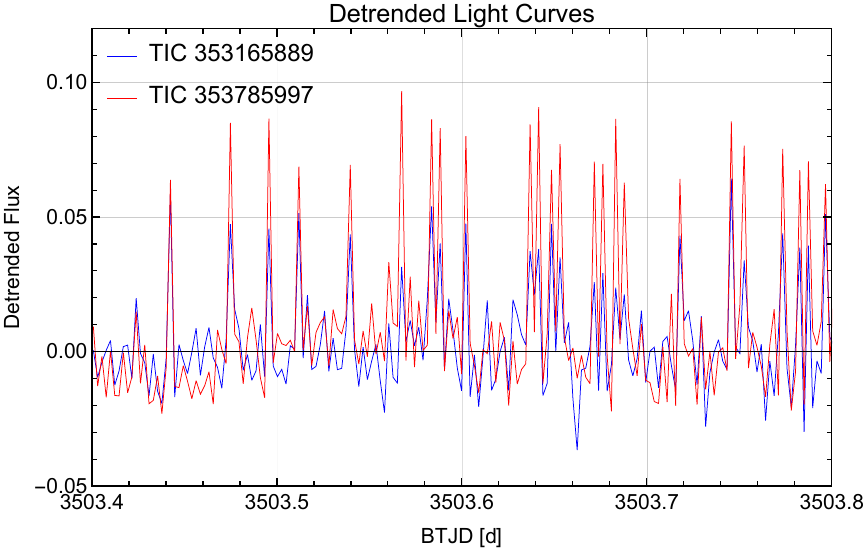}
  \caption{
Zoomed-in view of the correlated spike event in TIC~353165889 and
TIC~353785997 around BTJD~3503.
Both light curves are detrended using a 12\,h window.
Despite a sky separation of $771''$, the strongest excursions in this
interval occur at  the same 200 s {time bin}, demonstrating that the spike
is produced by a shared instrumental process rather than by independent
astrophysical variability.}
  \label{figure:fig4}
 \end{center}
\end{figure}

\section{Individual Verification of High-Significance Spikes}
\label{sec:global}

In this section we examine the five targets that warrant detailed inspection based on the statistical overview in Section~5 (see also Figs. \ref{fig:spike_lc_primary} and \ref{fig:spike_lc_summary}).  Two of these stars  (TIC~354057959 and TIC~353165889) host hard positive spikes inside the $\pm3.4$\,day BOAT arrival-time window.  The remaining three (TIC~10121249, TIC~9640566, and TIC~10121399) exhibit the most extreme excursions in the sample with $z\ge 10$ but outside their arrival windows. These objects form the natural set for vetting in the context of our technosignature search. Our assessment follows the diagnostic framework outlined in Section~6.1, which evaluates each spike in terms of single-{bin} morphology and same-bin coincidences among nearby stars.

\subsection{Diagnostic Framework}
\label{sec:simul_spike_framework}

\subsubsection{Motivation and classification problem}

In the BOAT--TESS sample, high-significance single-bin (200 s) excursions
may arise from three classes of phenomena:
(1) an ETI-related pulse tied to the BOAT illumination geometry,
(2) an astrophysical flare intrinsic to the star, and
(3) a short-lived instrumental artifact.
These three possibilities differ in two basic respects: their temporal morphology and their spatial coherence.
In the remainder of Section~6.1 we develop a diagnostic scheme based on
these two aspects and apply it to the five notable light curves identified in Section~5.

\subsubsection{Method 1: Morphology-based astrophysical vetting}

Astrophysical flares ordinarily evolve over timescales longer than a single
200\,s TESS exposure and produce multi-{bin} rise–decay structure
\citep{Davenport2014}. Ultra-short M-dwarf flares can occur within one {bin}
\citep{Aizawa2022}, but the five high-significance targets examined here have
absolute magnitudes $M_{T}\simeq 0$ or brighter
(based on TESS magnitudes combined with Gaia parallaxes; see
Table~\ref{tab:targets_major}), incompatible with M-dwarf hosts
($M_{T}\gtrsim 7$). Such ultra-short flares are therefore not expected in this subset.

  As discussed in Sec. 5.1, there are no consecutive hard positive excursions in any of the 58 light curves, including the five targets.

Taken together, the absolute magnitudes and the absence of sustained
multi-{bin} structure argue against an astrophysical-flare origin for all
five high-significance targets.

\subsubsection{Method 2: Coincidence and Timing Diagnostics}

The second diagnostic method examines whether a high-significance
single-{bin} excursion shows any spatial or temporal coherence across
multiple stars. Such coherence is not expected for astrophysical
variability or for an ETI-related pulse, whereas short-lived instrumental
disturbances can naturally generate same-{bin} anomalies in multiple
apertures.

\textit{{Bin}-level coincidence test.}
For each hard spike of interest we survey all QLP light curves within
$300''$ and identify neighboring events with $z\ge 5$ inside a
$\pm0.5$\,day window centered on the target spike.
A coincidence is defined as a neighboring event occurring in the same
200\,s exposure.
The probability that two independent events fall in the same {bin} within
a one-day interval is $\simeq 1/432$ per neighbor, so multi-star
coincidences are difficult to reconcile with independent astrophysical
variability.
In addition, TESS PRF spillover drops to only a few percent at separations
$\gtrsim 50''$ \citep{Feinstein2019}, making it unlikely that optical
contamination can reproduce multi-percent excursions at such distances.
These considerations make {bin}-level coincidences more consistent with
short-lived instrumental processes.

\textit{BOAT timing consistency.}
In the BOAT illumination geometry the pulse arrival time is a strong
function of the opening angle $\beta$; a difference
$\Delta\beta\simeq 0.03^\circ$ corresponds to a shift of $\Delta t_{\rm obs}\sim 50$ days.
Even a small separation of $\Delta\beta=50''$ implies a timing offset of
$\Delta t_{\rm obs}\sim 20$\,days, which is far larger than the $\pm 3.4$\,day BOAT window.
Same-{bin} synchrony among such stars is therefore not supported under an
ETI-related interpretation.

\textit{Interpretation across the three physical scenarios.}
Taken together, the coincidence test (including the BOAT timing relation) helps
discriminate among the three broad classes of explanations:
(i) an ETI-related pulse would be confined to a single star and would follow
the $\beta$-dependent arrival-time constraint,
(ii) astrophysical variability is also confined to one star and does not
produce inter-star coincidences, and
(iii) instrumental disturbances can generate same-{bin} excursions across
multiple apertures.
In Sections~6.2--6.4 we apply this framework to the five high-significance
targets in our sample.

We use the QLP aperture-photometry light curves, which provide
background-subtracted and systematics-mitigated fluxes suitable for
uniform comparisons across the nearby stars.
While the present analysis relies on these homogeneous products,
pixel-level information  can be valuable
for targeted follow-up of individual events
\citep[e.g.][]{Feinstein2019, Twicken2020}.
Such pixel-level diagnostics represent a natural direction for future
refinement of BOAT-related searches.

\begin{table*}
\centering
\caption{
Summary of representative hard-spike events in Sections~6.2–6.4.
For each TIC, we list one hard spike with its epoch (BTJD) and excursion level $z$.
“In-window” denotes events occurring within the BOAT arrival-time window.
The coincidence-test columns report whether a neighboring star within $300''$
shows a contemporaneous hard spike ($z\ge5$), together with the angular separation $\Delta$.
For TIC~354057959, two coincident neighboring stars are identified.
For TIC~353165889, only the strongest in-window event is shown.
}
\label{tab:event_summary}
\begin{tabular}{lcccccc}
\hline\hline
TIC ID
& Spike BTJD
& $z$
& In-window
& Coincident neighbor within $300''$
& $\Delta$ ($''$)
& Assessment \\
\hline
354057959
& 3508.406581
& 5.3
& Yes 
& Yes
& 56.2,~60.9
& Instrumental \\

353165889
& 3505.166961
& 5.9
& Yes 
& No
& ---
& Instrumental \\

10121249
& 3520.707430
& 10.8
& No
& Yes
& 50.2
& Instrumental \\

9640566
& 3494.191050
& 10.2
& No
& Yes
& 92.7
& Instrumental \\

10121399
& 3507.422834
& 10.1
& No
& Yes
& 58.1
& Instrumental \\
\hline
\end{tabular}
\end{table*}

\subsection{TIC 354057959}

As the first of the five targets examined in Sections~6.2--6.5,
TIC~354057959 (at $\beta=0.84865^\circ$) illustrates how the second diagnostic test outlined in
Section~6.1.3 is applied in practice, and serves as a reference example for how the results of the event-level vetting are condensed into the summary presented in Table~\ref{tab:event_summary}.  
Throughout this section we denote by $\Delta$ the angular separation
between the target and its neighboring stars.

This object hosts one hard spike inside the $\pm 3.4$\,day BOAT window
(Fig.~\ref{fig:spike_lc_primary}). 
The star is moderately bright ($T_{\rm mag}=11.8$) and shows limited overall
activity, with $N(|z|\ge 3.5)=14$ and 
$N(z\ge 5)=1$.

\subsubsection{In-window hard spike}

The in-window event occurs at BTJD~3508.406581 with $z=5.34$, close to the
raw value $z_{\rm raw}=5.36$, indicating that the 12\,h detrending has only a
minor effect.
The excursion lies entirely within a single stable segment (Segment~11), and
its full $\pm 0.5$\,day neighborhood is also contained within the same
segment.

\subsubsection{Nearby-star coincidence analysis}

We examined all QLP light curves within $\Delta \le 300\arcsec$ of the
target\add{, adopting the same search radius and {integration}-level criteria for all events summarized in Table~\ref{tab:event_summary}}.  
Among the 34 neighboring stars, exactly two show hard excursions
($z\ge5$) within the $\pm 0.5$\,day interval surrounding the event:
TIC~354057934 ($z=8.1$, $\Delta=60.9\arcsec$, $T_{\rm mag}=12.9$, $\beta=0.8658^\circ$) and
TIC~354057916 ($z=6.0$, $\Delta=56.2\arcsec$, $T_{\rm mag}=13.1$, $\beta=0.8642^\circ$).  
Both spikes occur in the same TESS exposure as the target spike, in contrast to
the expected BOAT-induced delay of $\Delta t_{\rm obs}\sim 20$ days.
No other hard excursions are found in this interval.

These angular separations correspond to $\sim2.5$--3 pixels, where PRF
spillover cannot generate multi-percent brightenings.
Optical contamination is therefore excluded as an explanation.
Under the assumption that spike timings are independent, the probability that
two neighboring hard spikes align with the target spike is of order
$(1/432)^2${; this estimate is quoted only as a heuristic indicator of rarity and is not used as a formal likelihood}.

\subsubsection{Implications for the BOAT search}

As noted in Section~6.1, the BOAT timing geometry does not predict
same-{bin} brightenings for stars with different $\beta$.  
The presence of two {exposure}-matched neighboring spikes therefore disfavors
an ETI-related interpretation within the BOAT-ring framework.  
Taken together, the synchrony, PRF-based contamination limits, and the
single-{bin} morphology strongly favor an instrumental origin for this
event.

\subsection{TIC 353165889}

TIC~353165889 ($T_{\rm mag}=12.9$, $\beta=0.84669^\circ$) exhibits one of the most complex
light-curve patterns in our sample, and therefore warrants a more extended discussion than the other targets. As shown in
Fig.~\ref{fig:spike_lc_primary}, Sector~80 contains a dense cluster of
positive excursions near BTJD~3503--3504, whereas Sector~81 shows a single
strong out-of-window spike. The latter event, a $z=7.66$ excursion at
BTJD~3528.839135, is the fourth-largest positive spike among all
58 BOAT-ring targets and provides a natural starting point for the
diagnostic analysis (not listed in Table 2). The raw value is nearly identical
($z_{\rm raw}=7.58$), illustrating the small impact of the 12\,h
detrending also for this event.

\subsubsection{A strong but out-of-window spike in Sector~81}
We first discuss the $z=7.66$ spike only to
clarify the broader systematic context.  
Inspecting all 22 QLP stars within $300''$, we find exactly one neighboring star,
TIC~353516097, showing a hard excursion at the {same} {200 s time bin}
($z=7.1$, $\Delta=87.7''$).  

This {bin}-level synchrony is confined to Sector~81, as TIC~353516097 shows no
corresponding activity in Sector~80.
Such sector-dependent coincidences are characteristic of short-lived
instrumental artifacts in TESS FFIs \citep{Davenport2016}.

\subsubsection{Initial inspection in Sector~80}

The strongest excursion in the arrival time window occurs at BTJD~3505.166961 with
$z=5.9$ (presented in Table~\ref{tab:event_summary}).  
A search among the 16 QLP stars within $300''$ of TIC~353165889 reveals no
hard ($z\ge 5$) coincidences within the same $\pm0.5$\,day interval
(Segment~8).  
Moreover, although the target shows seven in-window spikes with $z\ge 5$,
none of these events is accompanied by a contemporaneous hard spike
($z \ge 5$) in any other QLP star within $300''$
at the same 200\,s exposure.

These
results initially suggested that the Sector 80 spike might be
isolated. However, the dense cluster of positive excursions near
BTJD 3503–3504 motivated a broader investigation  beyond the minimal coincidence test
described in Section~6.1.3.

\subsubsection{Further study with TIC~353785997 and its environment}

To broaden the diagnostic search, we re-examined several dozen 
QLP light curves generated during an earlier quick-look
stage. TIC~353785997 ($T_{\rm mag}=11.3$;
RA $=288.97264^\circ$, DEC $=20.27675^\circ$), located
$\Delta=771''$ from TIC~353165889, shows a series of strong excursions around
BTJD~3503--3504 (see Fig.~3).\footnote{This star lies on the BOAT arrival ring
but does not pass the filters in either Section~3.4 or Section~3.5.}
When the two light curves are overplotted, their largest spikes are nearly
perfectly synchronized at the {bin} level (Fig.~\ref{figure:fig4}),
indicating a shared instrumental origin consistent with multi-aperture
coincidences reported in TESS systematics studies \citep{Jenkins2016}.

Using the strongest spike of TIC~353785997 ($z=11.58$ at
BTJD~3503.160028) as a reference, we examined all QLP sources within $300''$
(21 in total) and identified two additional objects,
TIC~353785989 ($\Delta\simeq4.1''$, $z=8.48$) and
TIC~353785962 ($\Delta\simeq40.8''$, $z=5.1$), exhibiting hard excursions
within the same $\pm0.5$\,day interval.
{These stars form a compact cluster of spike-affected apertures,}
and the synchronized $z=5.9$ excursion in TIC~353165889 is naturally
interpreted as a weaker imprint of the same short-lived localized systematic.

\subsubsection{CCD geometry and edge-related considerations}

In Sector~80, TIC~353165889 lies close to the outer boundary of Camera~2,
CCD~4 (see Fig.~2). Within the 58 BOAT-ring targets it is the second closest to
the CCD edge; the closest object, TIC~353166003 ($T_{\rm mag}=12.1$), exhibits a
relatively quiet light curve, suggesting that edge proximity is
associated with increased susceptibility to systematics but is not sufficient
by itself to produce spikes.

Column-aligned artifacts are a possible contributor, since some TESS
systematics propagate along CCD columns \citep{Jenkins2016}. However,
TIC~353165889 and TIC~353785997 are not column-aligned (relative offsets
$-32.4,+6.9$ pixels).
The neighbor with the strongest Sector~81 coincidence,
TIC~353516097, is much closer in the column direction (relative offsets
$+5.2,+10.3$ pixels), yet shows no corresponding activity in Sector~80.
These facts together suggest that the relevant systematics operate on spatial
scales larger than a single column, likely involving extended background
structures or sector-specific calibration features
\citep{Huang2020, Caldwell2020}.

\subsubsection{Summary}

TIC~353165889 exhibits different synchrony partners in different sectors.
In Sector~80 it correlates with TIC~353785997, TIC~353785989, and
TIC~353785962, while in Sector~81 it synchronizes with TIC~353516097.
Meanwhile, TIC~353785997 is quiet in Sector~81 ($z<4.45$).
This sector-dependent switching of co-excursion partners is incompatible
with any astrophysical or technosignature interpretation and instead reflects
spatially extended, short-lived instrumental systematics.

The behavior of TIC~353165889 is therefore best explained by CCD-edge
susceptibility and a spike-prone local environment, consistent with extended
detector anomalies reported in TESS FFI analyses.
{Accordingly, all in-window hard spikes for this target are attributed to instrumental effects.}

\subsection{Out-of-window hard spikes}

\add{
Three BOAT--TESS targets, TIC~10121249, TIC~9640566, and TIC~10121399,
host the three strongest hard spikes in the full sample (see Fig. \ref{fig:spike_lc_summary}).
All of these events occur outside the BOAT arrival-time window and 
serve as representative tests of the
\add{integration}-level coincidence diagnostics introduced in Section~6.1.
}

\add{
For each of these targets, the hard spike is a single-\add{integration}
excursion and is accompanied by a contemporaneous hard excursion
($z\ge5$) in at least one neighboring star within $\Delta \le 300''$,
as summarized in Table~\ref{tab:event_summary}.
}

\add{
Given the combination of out-of-window timing, single-\add{bin} morphology,
and clear \add{bin}-level coincidences, these events are most naturally explained by short-lived instrumental
systematics, following the criteria outlined in Section~6.1.
}

\add{
For completeness, we note that TIC~9640566 lies close to the Sector~81 boundary, and that TIC~10121249 is located at a distance of
$\sim8$\,kpc, exceeding the causal depth $r\cos\theta\sim5$kpc of the BOAT geometry.
}

\subsection{Summary of ETI-oriented vetting}

Among the 58 BOAT-ring targets, two stars (TIC~354057959 and
TIC~353165889) exhibit hard positive excursions inside the
$\pm 3.4$\,day BOAT window.  
{To place these events in broader context, we also examined the three stars
with the largest out-of-window excursions ($z\ge 10$), forming a set of
five targets for detailed vetting (see Table 2).}

Across these five targets, every hard positive spike is confined to a single 200\,s
exposure and lacks multi-{bin} duration.  
Each target spike  has at least one neighboring star showing a hard spike in the
same {exposure}, consistent with short-lived detector artifacts in TESS \citep{Davenport2016,Feinstein2019}.  
In nearly all cases the coincident neighbor lies within
{$\sim100''$}, while TIC~353165889 demonstrates that such systematics
can extend over separations of {up to $\sim800''$}.

Combining the two diagnostic tests (temporal morphology and
{bin}-level coincidences), none of the five examined cases shows
evidence supporting an ETI-related optical pulse{ within the BOAT timing
framework}.  
Instead, the results illustrate the practical value of narrow-window
spike vetting and coincidence analysis for identifying instrumental
false positives in TESS-based BOAT-ring searches.

\section{Summary and Discussion}

The hybrid ring strategy provides a geometrically motivated 
Schelling–point framework for SETI \citep{Seto2025}.  
By combining a precisely timed extragalactic transient with the 
high-precision distance to the Galactic center ($\sim0.5\%$), the 
arrival time depends strongly on the angular separation $\beta$.  
This offers a coordination scheme that does not assume any 
specific signaling format and is naturally compatible with 
continuous-survey instruments 
\citep[][]{Schelling1960,Wright2018,Tarter2001}. 

GRB\,221009A (BOAT) provides an exceptionally favorable reference for 
applying this geometry to real data.  
Because the relevant ring region was observed nearly  continuously by 
\textit{TESS} in Sectors~80 and~81, the predicted arrival-time window 
could be tested directly without dedicated follow-up.  
This is the first observational demonstration of the hybrid ring method, illustrating how a single burst can be used as a predictive, survey-based technosignature experiment.

A short artificial optical pulse, regardless of intrinsic duration, 
would appear in the 200\,s FFIs as a positive excursion confined to a single 200\,s exposure.
This asymmetry in the integrated photometry motivates focusing on large positive 
brightenings when identifying potential technosignature candidates.  
Guided by this, we selected the high-significance spikes in Section~5 for 
closer inspection:  
two stars (TIC~354057959 and TIC~353165889) show in-window 
($z\ge5$) excursions, and three additional stars 
(TIC~10121249, TIC~9640566, and TIC~10121399) exhibit the strongest 
out-of-window excursions ($z\ge10$).

The five targets analyzed in Section~6 provide a focused test of the hybrid-ring strategy under real survey conditions. While no evidence for intentional pulses is found within the parameter space explored here, several practical conclusions follow.

First, the coincidence test based on identical 200\,s exposures was uniformly decisive.
Every high-$z$ excursion—including the two in-window candidates—had at 
least one neighboring star showing a hard spike in the \emph{same} 
200\,s exposure, strongly supporting an instrumental origin.

Second, the spatial extent of these systematics was empirically measured.
Most coincidences occurred within $\sim100''$, yet TIC~353165889 showed
correlated behavior out to nearly $800''$, well beyond expectations from
simple PRF contamination.

Third, targets near CCD edges displayed increased susceptibility to 
spike-like anomalies.  While not deterministic, edge geometry offers 
useful diagnostic context.

Finally, the narrow $\pm3.4$\,day BOAT window restricted the number of 
high-$z$ events and  simplified the diagnostic procedure.

Taken together, these findings show that the hybrid-ring strategy is readily applicable to TESS survey photometry, and that coincidence tests based on identical 200\,s exposures provide an efficient filter for instrumental false
positives.
\add{To place these null results in a quantitative context, Appendix~B 
provides analytic upper limits on the transmitting power based on 
the observed photometric sensitivity of the QLP light curves analyzed here.
}

The present analysis is shaped by the basic properties of the TESS FFI data.  
With 200\,s integrations, any nanosecond to millisecond emission would be
recorded as a single integrated brightening, diluting its instantaneous 
peak flux and reducing the contrast against stochastic noise.  
In addition, broadband aperture photometry removes spectral information 
and mixes any narrowband or line-dominated pulse with the full passband, significantly reducing the effective signal-to-noise  ratio.
{These factors imply that only relatively high-fluence pulses would
be detectable as hard spikes in our standardized flux metric.}

These limitations arise from relying on widely accessible survey data 
rather than dedicated technosignature instruments, but the absence of 
confirmed ETI signals motivates continued use of all available datasets.  
Survey-oriented approaches therefore remain complementary to 
targeted searches.

Because the hybrid strategy is emission-agnostic, it can be applied 
across a wide range of pulse timescales and wavelengths.  
Future high-resolution, time-resolved facilities can similarly exploit
the same geometric coordination demonstrated here.
Other wide-field missions and upcoming surveys, such as PLATO or
Roman, could similarly leverage bright transients as anchors for
hybrid-ring searches, extending the BOAT-based proof of concept.

\add{
\begin{acknowledgments}
The author thanks the anonymous referee for a careful reading of the manuscript
and for constructive comments that helped improve the clarity and presentation
of this work.
This work made use of data from the TESS mission obtained from the Mikulski Archive
for Space Telescopes (MAST) at STScI \citep{mast_tess_ffi}.
This work has made use of data from the European Space Agency (ESA) mission Gaia,
processed by the Gaia Data Processing and Analysis Consortium (DPAC).
TESS focal-plane geometry and CCD-coordinate calculations employed the publicly
available \texttt{tess-point} tools developed by the MIT TESS team.
\end{acknowledgments}

\software{
Astropy \citep{astropy2013,astropy2018,astropy2022},
Matplotlib \citep{hunter2007},
NumPy \citep{vandewalt2011,harris2020},
SciPy \citep{jones2001},
pandas \citep{mckinney2010},
Lightkurve \citep{lightkurve2018}
}
}

\bibliography{refe}

\begin{thebibliography}{}
\expandafter\ifx\csname natexlab\endcsname\relax\def\natexlab#1{#1}\fi
\providecommand{\url}[1]{\href{#1}{#1}}
\providecommand{\dodoi}[1]{doi:~\href{http://doi.org/#1}{\nolinkurl{#1}}}
\providecommand{\doeprint}[1]{\href{http://ascl.net/#1}{\nolinkurl{http://ascl.net/#1}}}
\providecommand{\doarXiv}[1]{\href{https://arxiv.org/abs/#1}{\nolinkurl{https://arxiv.org/abs/#1}}}

\bibitem[{{Aizawa} {et~al.}(2022){Aizawa}, {Kawana}, \& ...}]{Aizawa2022}
{Aizawa}, M., {Kawana}, K., \& ... 2022, \pasj, 74, 1069,
  \dodoi{10.1093/pasj/psac056}

\bibitem[{{Astropy Collaboration}(2013)}]{astropy2013}
{Astropy Collaboration}. 2013, A\&A, 558, A33,
  \dodoi{10.1051/0004-6361/201322068}

\bibitem[{{Astropy Collaboration}(2018)}]{astropy2018}
---. 2018, AJ, 156, 123, \dodoi{10.3847/1538-3881/aabc4f}

\bibitem[{{Astropy Collaboration}(2022)}]{astropy2022}
---. 2022, ApJ, 935, 167, \dodoi{10.3847/1538-4357/ac7c74}

\bibitem[{Born \& Wolf(1999)}]{BornWolf1999}
Born, M., \& Wolf, E. 1999, Principles of Optics, 7th edn. (Cambridge
  University Press)

\bibitem[{{Burns} {et~al.}(2023){Burns}, {Svinkin}, \& ...}]{Burns2023}
{Burns}, E., {Svinkin}, D., \& ... 2023, \apjl, 946,
  \dodoi{10.3847/2041-8213/acc39c}

\bibitem[{{Cabrales} {et~al.}(2024){Cabrales}, {Davenport}, {Sheikh}, {Croft},
  {Siemion}, {Giles}, \& {Cody}}]{Cabrales2024a}
{Cabrales}, B., {Davenport}, J. R.~A., {Sheikh}, S., {et~al.} 2024, \aj, 167,
  101, \dodoi{10.3847/1538-3881/ad2064}

\bibitem[{Caldwell {et~al.}(2020)}]{Caldwell2020}
Caldwell, D.~A., {et~al.} 2020, Research Notes of the AAS, 4, 201,
  \dodoi{10.3847/2515-5172/abc9b3}

\bibitem[{Christiansen {et~al.}(2012)Christiansen, Jenkins, Caldwell,
  {et~al.}}]{Christiansen2012}
Christiansen, J.~L., Jenkins, J.~M., Caldwell, D.~A., {et~al.} 2012, PASP, 124,
  127

\bibitem[{{Davenport}(2016)}]{Davenport2016}
{Davenport}, J. R.~A. 2016, \apj, 829, 23, \dodoi{10.3847/0004-637X/829/1/23}

\bibitem[{{Davenport} {et~al.}(2022){Davenport}, {Cabrales}, {Sheikh}, {Croft},
  {Siemion}, {Giles}, \& {Cody}}]{Davenport2022}
{Davenport}, J. R.~A., {Cabrales}, B., {Sheikh}, S., {et~al.} 2022, \aj, 164,
  117, \dodoi{10.3847/1538-3881/ac82ea}

\bibitem[{{Davenport} {et~al.}(2014){Davenport}, {Hawley}, {Hebb},
  {Wisniewski}, {Kowalski}, {Johnson}, {Malatesta}, {Peraza}, {Keil},
  {Silverberg}, {Jansen}, {Scheffler}, {Berdis}, {Larsen}, \&
  {Hilton}}]{Davenport2014}
{Davenport}, J. R.~A., {Hawley}, S.~L., {Hebb}, L., {et~al.} 2014, \apj, 797,
  122, \dodoi{10.1088/0004-637X/797/2/122}

\bibitem[{{Drake}(1961)}]{Drake1961}
{Drake}, F.~D. 1961, Physics Today, 14, 40, \dodoi{10.1063/1.3057500}

\bibitem[{Feinstein {et~al.}(2019)}]{Feinstein2019}
Feinstein, A.~D., {et~al.} 2019, PASP, 131, 094502,
  \dodoi{10.1088/1538-3873/ab291c}

\bibitem[{{Gaia Collaboration} {et~al.}(2023){Gaia Collaboration}, Babusiaux,
  \& et~al.}]{Gaia2022}
{Gaia Collaboration}, Babusiaux, C., \& et~al. 2023, Astronomy \& Astrophysics,
  674, A4, \dodoi{10.1051/0004-6361/202243798}

\bibitem[{{GRAVITY Collaboration} {et~al.}(2021){GRAVITY Collaboration},
  {Abuter}, {Amorim}, {Baub{\"o}ck}, {Berger}, {Bonnet}, {Brandner},
  {Cl{\'e}net}, {Davies}, {de Zeeuw}, {Dexter}, {Dallilar}, {Drescher},
  {Eckart}, {Eisenhauer}, {F{\"o}rster Schreiber}, {Garcia}, {Gao}, {Gendron},
  {Genzel}, {Gillessen}, {Habibi}, {Haubois}, {Hei{\ss}el}, {Henning},
  {Hippler}, {Horrobin}, {Jim{\'e}nez-Rosales}, {Jochum}, {Jocou}, {Kaufer},
  {Kervella}, {Lacour}, {Lapeyr{\`e}re}, {Le Bouquin}, {L{\'e}na}, {Lutz},
  {Nowak}, {Ott}, {Paumard}, {Perraut}, {Perrin}, {Pfuhl}, {Rabien},
  {Rodr{\'\i}guez-Coira}, {Shangguan}, {Shimizu}, {Scheithauer}, {Stadler},
  {Straub}, {Straubmeier}, {Sturm}, {Tacconi}, {Vincent}, {von Fellenberg},
  {Waisberg}, {Widmann}, {Wieprecht}, {Wiezorrek}, {Woillez}, {Yazici},
  {Young}, \& {Zins}}]{Gravity2021}
{GRAVITY Collaboration}, {Abuter}, R., {Amorim}, A., {et~al.} 2021, \aap, 647,
  A59, \dodoi{10.1051/0004-6361/202040208}

\bibitem[{Harris {et~al.}(2020)Harris, Millman, van~der Walt,
  {et~al.}}]{harris2020}
Harris, C.~R., Millman, K.~J., van~der Walt, S.~J., {et~al.} 2020, Nature, 585,
  357, \dodoi{10.1038/s41586-020-2649-2}

\bibitem[{{Hippke}(2018)}]{2018JApA...39...73H}
{Hippke}, M. 2018, Journal of Astrophysics and Astronomy, 39, 73,
  \dodoi{10.1007/s12036-018-9566-x}

\bibitem[{Huang {et~al.}(2020)}]{Huang2020}
Huang, C.~X., {et~al.} 2020, Research Notes of the AAS, 4, 204,
  \dodoi{10.3847/2515-5172/abca2e}

\bibitem[{Hunter(2007)}]{hunter2007}
Hunter, J.~D. 2007, Computing in Science \& Engineering, 9, 90,
  \dodoi{10.1109/MCSE.2007.55}

\bibitem[{{Jenkins} {et~al.}(2016){Jenkins}, {Ricker}, {Seager}, {Latham},
  {Vanderspek}, {Winn}, {Twicken}, {Caldwell}, {Shao}, {Quinn}, {McLeod},
  {Villasenor}, {Klaus}, \& {Vezie}}]{Jenkins2016}
{Jenkins}, J.~M., {Ricker}, G.~R., {Seager}, S., {et~al.} 2016, in Space
  Telescopes and Instrumentation 2016: Optical, Infrared, and Millimeter Wave,
  Vol. 9913, 99133E, \dodoi{10.1117/12.2233418}

\bibitem[{{Lemarchand}(1994)}]{1994Ap&SS.214..209L}
{Lemarchand}, G.~A. 1994, \apss, 214, 209, \dodoi{10.1007/BF00982337}

\bibitem[{{Lightkurve Collaboration}(2018)}]{lightkurve2018}
{Lightkurve Collaboration}. 2018, Astrophysics Source Code Library

\bibitem[{{Makovetskii}(1980)}]{1980Icar...41..178M}
{Makovetskii}, P.~V. 1980, \icarus, 41, 178,
  \dodoi{10.1016/0019-1035(80)90002-0}

\bibitem[{McKinney(2010)}]{mckinney2010}
McKinney, W. 2010, in Proceedings of the 9th Python in Science Conference,
  56--61

\bibitem[{{McLaughlin}(1977)}]{1977Icar...32..464M}
{McLaughlin}, W.~I. 1977, \icarus, 32, 464,
  \dodoi{10.1016/0019-1035(77)90019-7}

\bibitem[{{Nilipour} {et~al.}(2023){Nilipour}, {Davenport}, \&
  ...}]{Nilipour2023}
{Nilipour}, A., {Davenport}, J., \& ... 2023, \aj, 166,
  \dodoi{10.3847/1538-3881/acde79}

\bibitem[{Ricker {et~al.}(2015)Ricker, Winn, Vanderspek, \&
  et~al.}]{Ricker2015}
Ricker, G.~R., Winn, J.~N., Vanderspek, R., \& et~al. 2015, Journal of
  Astronomical Telescopes, Instruments, and Systems, 1, 014003,
  \dodoi{10.1117/1.JATIS.1.1.014003}

\bibitem[{Schelling(1960)}]{Schelling1960}
Schelling, T.~C. 1960, The Strategy of Conflict (Harvard University Press)

\bibitem[{{Seto}(2019)}]{Seto2019}
{Seto}, N. 2019, \apjl, 875, \dodoi{10.3847/2041-8213/ab133a}

\bibitem[{{Seto}(2021)}]{Seto2020}
---. 2021, \apj, 917, \dodoi{10.3847/1538-4357/ac0c7b}

\bibitem[{{Seto}(2024)}]{Seto2024}
---. 2024, \apj, 964, \dodoi{10.3847/1538-4357/ad2a48}

\bibitem[{{Seto}(2025)}]{Seto2025}
---. 2025, \apj, 994, \dodoi{10.3847/1538-4357/ae06a8}

\bibitem[{{Stassun} {et~al.}(2019){Stassun}, {Oelkers}, {Paegert}, {Torres},
  {Pepper}, {De Lee}, {Collins}, {Latham}, {Muirhead}, {Chittidi},
  {Rojas-Ayala}, {Fleming}, {Rose}, {Tenenbaum}, {Ting}, {Kane}, {Barclay},
  {Bean}, {Brassuer}, {Charbonneau}, {Ge}, {Lissauer}, {Mann}, {McLean},
  {Mullally}, {Narita}, {Plavchan}, {Ricker}, {Sasselov}, {Seager}, {Sharma},
  {Shiao}, {Sozzetti}, {Stello}, {Vanderspek}, {Wallace}, \&
  {Winn}}]{Stassun2019}
{Stassun}, K.~G., {Oelkers}, R.~J., {Paegert}, M., {et~al.} 2019, \aj, 158,
  138, \dodoi{10.3847/1538-3881/ab3467}

\bibitem[{{Tarter}(2001)}]{Tarter2001}
{Tarter}, J. 2001, \araa, 39, 511, \dodoi{10.1146/annurev.astro.39.1.511}

\bibitem[{{TESS Team}(2022)}]{mast_tess_ffi}
{TESS Team}. 2022, TESS Calibrated Full Frame Images: All Sectors,  STScI/MAST,
  \dodoi{10.17909/0CP4-2J79}

\bibitem[{Twicken {et~al.}(2020)Twicken, Caldwell, Jenkins,
  {et~al.}}]{Twicken2020}
Twicken, J.~D., Caldwell, D.~A., Jenkins, J.~M., {et~al.} 2020, TESS Science
  Data Products Description Document, Tech. Rep. NASA/TM--20205008729, NASA
  Ames Research Center.
\newblock
  \url{https://archive.stsci.edu/missions/tess/doc/EXP-TESS-ARC-ICD-TM-0014-Rev-F.pdf}

\bibitem[{van~der Walt {et~al.}(2011)van~der Walt, Colbert, \&
  Varoquaux}]{vandewalt2011}
van~der Walt, S., Colbert, S.~C., \& Varoquaux, G. 2011, Computing in Science
  \& Engineering, 13, 22, \dodoi{10.1109/MCSE.2011.37}

\bibitem[{{Virtanen} {et~al.}(2020){Virtanen}, {Gommers}, {Oliphant},
  {Haberland}, {Reddy}, {Cournapeau}, {Burovski}, {Peterson}, {Weckesser},
  {Bright}, {van der Walt}, {Brett}, {Wilson}, {Millman}, {Mayorov}, {Nelson},
  {Jones}, {Kern}, {Larson}, {Carey}, {Polat}, {Feng}, {Moore}, {VanderPlas},
  {Laxalde}, {Perktold}, {Cimrman}, {Henriksen}, {Quintero}, {Harris},
  {Archibald}, {Ribeiro}, {Pedregosa}, {van Mulbregt}, \& {SciPy 1. 0
  Contributors}}]{jones2001}
{Virtanen}, P., {Gommers}, R., {Oliphant}, T.~E., {et~al.} 2020, Nature
  Medicine, 17, 261, \dodoi{10.1038/s41592-019-0686-2}

\bibitem[{{Wright}(2018)}]{Wright2018}
{Wright}, J.~T. 2018, in Handbook of Exoplanets, ed. H.~J. {Deeg} \& J.~A.
  {Belmonte}, 186, \dodoi{10.1007/978-3-319-55333-7_186}

\bibitem[{{Wright} {et~al.}(2018){Wright}, {Kanodia}, \& {Lubar}}]{Wright2018b}
{Wright}, J.~T., {Kanodia}, S., \& {Lubar}, E. 2018, \aj, 156, 260,
  \dodoi{10.3847/1538-3881/aae099}

\end{thebibliography}

\appendix
\section{Supplementary Tables}
\label{app:tables}

\startlongtable
\begin{deluxetable*}{rcccccccc}
\tablecaption{Basic properties of the final 58 QLP targets.
The coordinates (RA, Dec, $\beta$, $t_{\mathrm{obs}}$) are propagated with
proper-motion corrections to the mid-epoch of our TESS observations
(2024 July 14).  The last column gives the crowding category based on Gaia
neighbors: P (“pristine”) indicates no Gaia neighbor within $42\arcsec$ with
$\Delta G < 2.5$\,mag, while R (“restricted”) indicates at least one such
neighbor within $42\arcsec$ but none within $21\arcsec$. \add{The astrometric parameters (RA, Dec, parallax, and $T_{\rm eff}$) are taken from
Gaia DR3 \citep{Gaia2022}, while the TESS magnitude $T_{\rm mag}$ and target identifiers are from the
TESS Input Catalog \citep{Stassun2019} and QLP products \citep{Huang2020}.  The derived quantities ($\beta$ and
$t_{\rm obs}$) are computed in this work following the hybrid-ring geometry
of \citet{Seto2025}.}\label{tab:targets_major}}
\tablehead{
  \colhead{TIC} &
  \colhead{RA [deg]} &
  \colhead{Dec [deg]} &
  \colhead{$\beta$ [deg]} &
  \colhead{$t_{\rm obs}$ (BTJD)} &
  \colhead{$T_{\rm mag}$} &
  \colhead{Parallax  $\varpi$ [mas]} &
  \colhead{$T_{\rm eff}$ [K]}
}
\startdata
9419445 & 287.39776 & 19.98737 & 0.84277 & 3501.774 & 13.2 & 0.823 & 6787 & R \\
9419767 & 287.37792 & 19.94155 & 0.85072 & 3513.902 & 13.3 & 0.003 &  & R \\
9419783 & 287.37149 & 19.93957 & 0.85627 & 3522.437 & 12.9 & 0.192 & 4638 & R \\
9420241 & 287.37477 & 19.87037 & 0.84268 & 3501.644 & 11.4 & 2.467 & 6166 & P \\
9420675 & 287.35384 & 19.81250 & 0.85783 & 3524.845 & 12.9 & 0.995 & 6641 & R \\
9421079 & 287.37289 & 19.76118 & 0.83924 & 3496.422 & 13.4 & 0.103 &  & R \\
9423012 & 287.41949 & 19.51312 & 0.83739 & 3493.638 & 12.5 & 0.292 & 4821 & R \\
9423208 & 287.41392 & 19.48756 & 0.85069 & 3513.856 & 12.3 & 1.627 & 4897 & R \\
9633752 & 287.52687 & 19.30818 & 0.83653 & 3492.331 & 13.2 & 0.110 & 4837 & R \\
9633769 & 287.51963 & 19.31056 & 0.84089 & 3498.922 & 11.3 & 0.899 & 4747 & R \\
9634085 & 287.48029 & 19.35206 & 0.85070 & 3513.867 & 10.7 & 0.723 & 7999 & R \\
9634287 & 287.47479 & 19.37753 & 0.84289 & 3501.961 & 11.2 & 0.185 &  & P \\
9634394 & 287.46396 & 19.39210 & 0.84521 & 3505.493 & 13.1 & 0.116 &  & P \\
9639390 & 287.42759 & 20.04839 & 0.83361 & 3487.948 & 10.4 & 0.493 & 4695 & R \\
9639927 & 287.42986 & 20.11927 & 0.85749 & 3524.316 & 12.8 & 1.519 & 6342 & R \\
9640566 & 287.48369 & 20.20429 & 0.85099 & 3514.319 & 10.4 & 0.485 & 4789 & P \\
9640619 & 287.50668 & 20.21244 & 0.83666 & 3492.535 & 12.8 & 0.393 &  & R \\
9641063 & 287.53936 & 20.27344 & 0.84516 & 3505.416 & 13.3 & 0.215 & 4527 & R \\
10121163 & 287.76774 & 19.06593 & 0.84858 & 3510.632 & 11.8 & 0.616 & 5084 & R \\
10121249 & 287.77755 & 19.07617 & 0.83494 & 3489.940 & 11.6 & 0.376 &  & P \\
10121399 & 287.73799 & 19.09543 & 0.84039 & 3498.174 & 12.3 & 0.115 & 3806 & R \\
10121467 & 287.70107 & 19.10408 & 0.85462 & 3519.901 & 13.0 & 1.811 & 5866 & R \\
10449623 & 287.84848 & 19.03591 & 0.83543 & 3490.685 & 11.5 & 0.248 &  & R \\
10449826 & 287.88191 & 19.01076 & 0.84377 & 3503.293 & 11.1 & 0.234 & 3821 & R \\
351969625 & 287.99147 & 18.96811 & 0.84552 & 3505.954 & 11.8 & 0.862 &  & P \\
351969767 & 288.05609 & 18.94848 & 0.84806 & 3509.834 & 12.3 & 0.259 & 4714 & R \\
352086554 & 288.27786 & 18.92396 & 0.84954 & 3512.103 & 13.4 & 0.793 &  & R \\
352086582 & 288.18778 & 18.92866 & 0.84785 & 3509.519 & 11.3 & 0.357 & 3861 & P \\
352461926 & 288.38586 & 18.93261 & 0.84856 & 3510.596 & 12.1 & 1.917 & 5686 & P \\
352586364 & 288.45007 & 18.93842 & 0.85314 & 3517.612 & 11.5 & 0.567 &  & R \\
353015292 & 288.67550 & 19.01021 & 0.85599 & 3522.003 & 11.8 & 0.876 &  & R \\
353015529 & 288.62576 & 18.99416 & 0.85048 & 3513.530 & 12.7 & 0.259 & 4616 & R \\
353015772 & 288.58956 & 18.97805 & 0.85240 & 3516.487 & 13.2 & 1.170 & 5167 & R \\
353156178 & 288.71673 & 19.02818 & 0.85863 & 3526.087 & 12.3 & 0.181 &  & P \\
353165889 & 288.82266 & 20.43832 & 0.84660 & 3507.608 & 12.4 & 0.305 &  & P \\
353166003 & 288.77470 & 20.45525 & 0.83327 & 3487.430 & 12.1 & 0.798 & 5136 & P \\
353516270 & 288.85980 & 20.40830 & 0.84590 & 3506.538 & 11.1 & 1.842 &  & P \\
353516526 & 288.91786 & 20.36597 & 0.85301 & 3517.421 & 12.7 & 0.554 & 4708 & R \\
353516633 & 288.91567 & 20.34972 & 0.84033 & 3498.073 & 12.7 & 0.312 & 4520 & P \\
353524448 & 288.91651 & 19.18511 & 0.85079 & 3514.008 & 11.8 & 0.366 &  & P \\
353524786 & 288.86531 & 19.13348 & 0.85461 & 3519.882 & 12.0 & 0.357 & 4761 & R \\
353663443 & 289.00557 & 19.31832 & 0.83349 & 3487.764 & 11.3 & 0.942 & 4841 & P \\
353663626 & 289.04946 & 19.34096 & 0.85674 & 3523.167 & 11.3 & 1.941 & 6699 & P \\
353785185 & 289.06522 & 20.16326 & 0.84749 & 3508.963 & 12.6 & 0.197 & 4015 & R \\
353785479 & 289.04703 & 20.20432 & 0.85226 & 3516.270 & 11.1 & 2.364 &  & P \\
353785715 & 289.03076 & 20.23495 & 0.85519 & 3520.767 & 13.0 & 0.442 & 4877 & R \\
354057233 & 289.08301 & 20.14988 & 0.85643 & 3522.684 & 12.5 & 0.213 & 4687 & P \\
354057707 & 289.09960 & 20.08515 & 0.84464 & 3504.623 & 13.4 & 0.366 & 5561 & R \\
354057959 & 289.11795 & 20.04961 & 0.84856 & 3510.590 & 11.8 & 0.630 &  & R \\
354058108 & 289.12285 & 20.02889 & 0.84648 & 3507.427 & 13.2 & 0.351 & 5802 & R \\
354058157 & 289.12976 & 20.02058 & 0.85024 & 3513.169 & 13.0 & 1.034 & 6032 & R \\
354059147 & 289.15736 & 19.88338 & 0.84702 & 3508.242 & 13.2 & 0.173 &  & R \\
354059308 & 289.15058 & 19.85965 & 0.83797 & 3494.515 & 12.3 & 1.320 &  & R \\
354059504 & 289.17109 & 19.83135 & 0.85487 & 3520.277 & 11.8 & 0.425 &  & P \\
354060547 & 289.16686 & 19.68163 & 0.85427 & 3519.349 & 13.2 & 0.033 & 4726 & P \\
354061708 & 289.12085 & 19.51570 & 0.84660 & 3507.607 & 12.5 & 2.742 & 5443 & R \\
354062527 & 289.08830 & 19.40605 & 0.85860 & 3526.036 & 10.0 & 2.858 & 7963 & P \\
384964535 & 288.10623 & 18.93010 & 0.85645 & 3522.710 & 12.1 & 1.270 &  & R \\
\enddata
\end{deluxetable*}

\section{Required transmitting power}
\label{app:power}

For each BOAT--TESS target, we estimate the  transmitting power required
for a detectable optical brightening in the TESS light curve.
These estimates are then interpreted as observer-side upper limits, in the
sense that any transmitter exceeding the derived power would have produced
a detectable signal in our data.

\subsection{Stellar flux and detection threshold}
\label{app:power_sigma}

Throughout this Appendix, the index $i=1,\dots,58$ labels the individual
BOAT--TESS targets that passed all selection criteria.

For each target $i$, we define $\sigma_i$ as the time median of the robust
scatter $\sigma_{\rm robust}(t)$ defined in Eq.~(\ref{sigm}), evaluated over
the intersection of the $\pm3.4$~day BOAT arrival window and the set of observing segments common to all targets (Table~1).
This quantity characterizes the effective fractional noise level relevant for
single-bin excursions in the search window.

Across the 58 targets, the median value of $\sigma_i$ is
$\simeq5.5\times10^{-3}$, with the central 68\% of the sample (16th-84th percentiles) spanning
$3.0\times10^{-3}$ to $1.2\times10^{-2}$.
For each target, using the maximum in-window scatter instead of the median increases
the inferred value of $\sigma_i$ by at most a factor of $\sim2.3$ and does
not qualitatively affect our discussion below.

For a target with TESS magnitude $T_{{\rm mag},i}$, the observed stellar flux in
the TESS band is
\begin{equation}
F_{\star,i} = F_{T,0}\,10^{-0.4 T_{{\rm mag},i}},
\end{equation}
where $F_{T,0}=1.51\times10^{-5}\ {\rm erg\,s^{-1}\,cm^{-2}}$ is the
band-integrated TESS zero point \citep{Ricker2015}.

We define a detectable signal in terms of a fractional brightening relative to
the stellar flux.
The required signal flux is
\begin{equation}
F_{{\rm sig},i} = n_{\rm det}\,\sigma_i\,F_{\star,i},
\end{equation}
with a fixed detection factor $n_{\rm det}=5$ in  this appendix.
This definition ensures that the detection threshold is evaluated consistently
over the same arrival-time window used throughout the spike analysis.

\subsection{Isotropic-equivalent power}
\label{app:power_iso}

Given the distance $d_i$ inferred from the Gaia parallax $\varpi_i$ for target $i$, the
isotropic-equivalent power required to produce the detectable signal flux
$F_{{\rm sig},i}$ is
\begin{equation}
P_{{\rm iso},i} = 4\pi d_i^{\,2}\,F_{{\rm sig},i}.
\end{equation}

\subsection{Beamed transmission and fiducial assumptions}
\label{app:power_beam}

To translate the isotropic-equivalent bound into a required transmitting power, we consider beamed emission.
Assuming diffraction-limited radiation from a circular transmitting aperture
of diameter $D_{\rm tx}$ at wavelength $\lambda$, the beam solid angle is \citep{BornWolf1999}
\begin{equation}
\Omega \simeq \pi \left(1.22\,\frac{\lambda}{D_{\rm tx}}\right)^2 .
\end{equation}
Under continuous emission, the required transmitting power for target $i$ is
reduced relative to the isotropic-equivalent value by the geometric factor
$\Omega/4\pi$, yielding
\begin{equation}
P_{{\rm req},i} = P_{{\rm iso},i}\,\frac{\Omega}{4\pi}.
\end{equation}

Throughout the following, we adopt a fiducial set of parameters,
\begin{equation}
\lambda = 600~{\rm nm}, \qquad
D_{\rm tx} = 30~{\rm m},
\end{equation}
and assume that the signal is present continuously over the effective
integration time of a single TESS data point (200~s).
Under these assumptions, the required transmitting power
$P_{{\rm req},i}^{\rm fid}$ for each target is uniquely determined by
$P_{{\rm iso},i}$.

\subsection{Scaling with wavelength, aperture, and signal duration}
\label{app:power_scaling}

For different choices of wavelength $\lambda$, transmitting aperture
$D_{\rm tx}$, or for a signal present only for a shorter duration
$t_{\rm sig}\le 200~{\rm s}$, the required transmitting power for target $i$
scales as
\begin{equation}
P_{{\rm req},i}
=
P_{{\rm req},i}^{\rm fid}
\left(\frac{\lambda}{600~{\rm nm}}\right)^2
\left(\frac{30~{\rm m}}{D_{\rm tx}}\right)^2
\left(\frac{200~{\rm s}}{t_{\rm sig}}\right),
\label{eq:power_scaling_factor}
\end{equation}
where $P_{{\rm req},i}^{\rm fid}$ denotes the required power under the fiducial
assumptions defined above.

This expression shows that longer wavelengths, smaller transmitting apertures,
or shorter signal durations systematically increase the required power.
For example, a signal present for only $20~{\rm s}$ requires a factor of ten
higher power than the fiducial continuous-emission case.

\subsection{Distribution of required transmitting power}
\label{app:power_distribution}

Applying the fiducial assumptions described above to the set
$(T_{{\rm mag},i}, \varpi_i, \sigma_i)$ for all 58 BOAT--TESS targets, we obtain a well-defined distribution of the required transmitting power
$P_{{\rm req},i}^{\rm fid}$ under the fiducial assumptions adopted here.

Under the fiducial assumptions, the median required transmitting power across
the sample is $\simeq 53~{\rm GW}$.
The central 68\% of targets (16th--84th percentiles) span a range from
$\simeq 4.5~{\rm GW}$ to $\simeq 260~{\rm GW}$, indicating substantial
source-to-source variation driven primarily by differences in distance,
stellar brightness, and photometric noise.

As a representative best-case example, the most favorable target in the sample
(TIC~354061708) requires a transmitting power of order $1~{\rm GW}$ under the
same fiducial assumptions, illustrating that the lower end of the distribution
can reach the gigawatt scale for nearby and relatively bright sources.

These values quantify the observer-side upper limits implied by the null result
under the assumed transmission geometry.

\subsection{Comment on interstellar extinction}
The upper limits on the transmitting powers derived above would increase if explicit corrections for interstellar extinction were applied.

Gaia DR3 provides the extinction proxy \texttt{azero\_gspphot} $A_0$ for 37 of the
58 BOAT--TESS targets \citep{Gaia2022}.
Overall, the inferred extinction displays a clear negative correlation with
parallax, consistent with dust-dominated sight lines.
Targets with relatively large parallaxes ($\varpi \gtrsim 1.6$~mas) typically
exhibit modest extinction ($A_0 \lesssim 1.3$), whereas more distant targets
($\varpi \lesssim 0.16$~mas) show a broader range,
$A_0 \simeq 1.5$--5.2, with a substantial fraction lacking reliable estimates.

The BOAT geometry preferentially selects directions close to the Galactic disk,
where the stellar surface density is high but interstellar extinction is enhanced. This
reinforces a common motivation for technosignature searches at longer
wavelengths, where dust attenuation is substantially reduced \citep[e.g.,][]{2018JApA...39...73H}.

\end{document}